\documentclass[twocolumn,preprintnumbers,pra,superscriptaddress,showkeys]{revtex4}

\usepackage{amsmath,amssymb,amsfonts,graphicx,float,times,psfrag}
\usepackage[pdftex]{color}
\usepackage{amsmath,bm}
\usepackage[colorlinks, linkcolor=blue, citecolor=blue,  breaklinks=true]{hyperref}
\usepackage{mathtools,amsfonts,mathptmx}
\usepackage[utf8]{inputenc}
\usepackage[T1]{fontenc}
\usepackage{xcolor}
\usepackage{braket}
\usepackage[titletoc,title]{appendix}
\usepackage[nameinlink,capitalize]{cleveref}

\usepackage[shortlabels]{enumitem}
\begin{document}
\title{Noise-tolerant tripartite entanglement and quantum coherence via saturation effects} 

\author{P. Djorwé}
\email{djorwepp@gmail.com}
\affiliation{Department of Physics, Faculty of Science, 
University of Ngaoundere, P.O. Box 454, Ngaoundere, Cameroon}
\affiliation{Stellenbosch Institute for Advanced Study (STIAS), Wallenberg Research Centre at Stellenbosch University, Stellenbosch 7600, South Africa}

\author{J.-X. Peng}
\email{JiaXinPeng@ntu.edu.cn}
\affiliation{School of Physical Science and Technology, Nantong University, Nantong, 226019, People’s Republic of China}

\author{S. Adbel-Khalek}
\email{JiaXinPeng@ntu.edu.cn}
\affiliation{Department of Mathematics and Statistics, College of Science, Taif 21944, Saudi Arabia}

\author{A.-H. Abdel-Aty}
\email{amabdelaty@ub.edu.sa}
\affiliation{Department of Physics, College of Sciences, University of Bisha, Bisha 61922, Saudi Arabia}


\begin{abstract}
Engineering quantum resources that survive against environmental temperature is of great interest for modern quantum technologies. However, it is a tricky task to synthetize such quantum states. Here, we propose a scheme to generate highly resilient tripartite entanglement and quantum coherence against thermal fluctuations. Our benchmark model consists of a mechanical resonator driven by two electromagnetic fields, which are optically coupled. A modulated photon hopping $J$ captures the optical coupling, and each optical cavity hosts saturable gain or loss. When the saturable gain/loss are off, we observe a slightly enhancement of both tripartite entanglement and quantum coherence for an appropriate tuning of the phase modulation. When the saturation effects are turned on, we observe a significant enhancement of the tripartite entanglement, up to one order of magnitude, together with a moderate improvement of the quantum coherence. More interestingly, our results show that the threshold thermal phonon mumber for preserving tripartite entanglement in our proposal has been postponed up to two order of magnitude stronger than when the saturation effects are not accounted. The inclusion of saturable gain/loss in our proposal induces noise-tolerant quantum resources, and may lead to room temperature quantum applications such as quantum information processing, and quantum computional tasks. Our findings are quite general, and suggest saturation nonlinear effects as a tool for engineering thermal-immune quantum correlations.          
\end{abstract}

\maketitle
\date{\today}

\section{Introduction}\label{Intro}

Quantum coherence and quantum entanglement can be used to characterize quantum properties of a compound system. These two quantities constitute vital quantum resources for modern quantum technologies such as quantum information processing \cite{Wendin2017,Flamini2018,Slussarenko2019}, quantum communication \cite{Pittaluga2025,Li2025}, and quantum computational tasks \cite{Larocca2025,Proctor2025,Maring2024}. Quantum coherence deepens our understanding of the boundaries between classical and quantum worlds, and not only constitutes necessary condition for generating entanglement \cite{Cheng2015} but helps for optimizing and manipulating quantum correlations \cite{Bemani2019,Emale.2025,Rostand2025,Purdy2017}. Therefore, it is interesting to investigate on the common features exhibited by quantum coherence and quantum entanglement, and to simultaneously enhance them for further sophisticated technological applications.

Quantum coherence, as a basic and necessary condition to study quantum correlations has attracted increasing attention and great efforts
have been devoted for its quantification. Among the pioneering works aiming to quantify the relationship between quantum coherence and quantum correlations, one can mention \cite{Cheng2015,Zhu2017}. These works have laid down the fundamental relation between quantum coherence and quantum correlations, i.e., any entanglement measure of bipartite pure states is the minimum of a suitable coherence measure over product bases. Such an investigation has been extended later on to tripartite system \cite{Dong2022}, where the authors have shown that coherence and
quantum correlations are intrinsically related and can be converted to one another. Beside these fundamental investigations, a metric to quantify quantum coherence of Gaussian state in infinite-dimensional bosonic system based on the relative entropy has been established \cite{Zhang2016,Xu2016}. This has led to the quantification of quantum coherence in plethora of physical systems including, optomechanical structures \cite{Zheng2016,Peng2024}, magnomechanical systems \cite{Peng2024}, Josephson junctions \cite{Frwis2018}, and Bose-Einstein condensates \cite{Donley2002}. Moreover, experimental demonstrations of macroscopic quantum coherence have been reported \cite{Marquardt2007,Kang2021}, which may constitute steps towards practical applications.

Quantum entanglement captures the inseparability features of quantum correlations shared by distant entities. Bipartite entanglement that results from the interaction between two systems has been widely investigated  in optomechanical structures \cite{Zhai2023,Wu2023,Rost2024} and related  systems \cite{Fan2024,Hussain2022}. For proper applications, multipartite entanglement is required as it allows interaction between several parts of a system. Such multipartite entanglement is crucial for quantum computational tasks \cite{Tsukanov2011,Puri.2025}, for sensitivity improvement of sensing schemes \cite{Li2021, Dj2024, Xia2023,Tang2023,Djorwe2019}, and for quantum metrology \cite{Pezz2018,Bai2019}. However, engineering multipartite entanglement is not an easy task since multi interactions induce dark mode effect, destructive interferences, nonlinearities, and more dissipative channels which all constitute severe drawback for entanglement generation. Therefore, sophisticated approaches have to be established in order to synthetize multipartite entanglement. For instance, dark mode control has been used in \cite{Lai2022,Nori2022} to enhance entanglement in optomechanical systems. In the dark mode breaking regime, the authors have synthetized optomechanical tripartite entanglement that resists twice to thermal fluctuations \cite{Lai2022} and up to three order of magnitude \cite{Nori2022} more than that in the unbreaking regime. Such a thermal management of entanglement is assisted by synthetic gauge field, and is an interesting tool for achieving noise-tolerant tripartite entanglement. Control of both macroscopic quantum coherence and tripartite entanglement in a magnomechanical system has been realized in \cite{Qiu2022}. This control is done by simultaneously driving the system by a magnetic and a microwave field pumps. It has been shown that tripartite entanglement can only be realized in a small range of parameters, while the parameter's range for controlling quantum coherence is larger. In \cite{Liu2025,Jiao2025}, a squeezing-phase-controlled quantum noise has been used to selectively generate and manipulate tripartite entanglement and asymmetric EPR steering in a nonlinear whispering gallery mode. By adjusting the squeezing parameters, the authors have engineered various types of entanglement and EPR-steering states which are robust against environmental thermal noises. Moreover, nonreciprocal tripartite entanglement was induced in  magnon optomechanical system via Barnett effect in \cite{Ge2025,Lu2025}. Furthermore, nonlinearities have been used to synthetize tripartite entanglement in an exciton-optomechanical in \cite{Cai2025}, and to induce nonreciprocal tripartite entanglement in a cavity-magnon optomechanics \cite{Chen2023}. These nonlinear effects have led to a robust entanglement against thermal baths. 

Owing to these races towards thermal management of tripartite entanglement, it is crucial to come up with new approaches to engineer further noise-tolerant quantum states which stand as a bottleneck and  core of current quantum technologies. Therefore, we propose a scheme based on nonlinear saturation gain/loss which has been recently used to generate nonlinear exceptional points and to reduce noise amplification around these non-hermitian singularities \cite{Bai2022,Bai2023,Bai2024a}. Our benchmark system is a mechanical resonator driven by two blue-detuned electromagnetic fields that are optically coupled. The optical cavities host either saturable gain or loss, and the photon hopping rate capturing the optical coupling is phase modulated under a synthetic magnetism. Our findings state that i) both quantum coherence and tripartite entanglement are enhanced as the optical coupling increases for specific values of the phase modulation, ii) the synthetized entanglement is up to one order of magnitude  greater than what is achieved without the saturable gain/loss, and iii) the generated tripartite entanglement is up to two order of magnitude robust against thermal fluctuations than their counterparts generated in the absence of saturation effects. These findings shed light on nonlinear saturable gain/loss as tool for the generation of noise-tolerant quantum correlations, which are useful for thermal management of quantum resources in modern quantum technologies.    

The rest of the work is organized as follows. The \autoref{sec:mod} describes the model and the related dynamical equations. In \autoref{sec:metr}, we present the metrics used to quantify quantum coherence and tripartite entanglement. The effects of the optical coupling are investigated in \autoref{sec:coupl}, while those for the saturable effects are carried out in \autoref{sec:sat}. We conclude our work in \autoref{sec:concl} where we summarize our findings.     

\section{Model and dynamical equations}\label{sec:mod}
\begin{figure}[tbh]
\begin{center}
  \resizebox{0.45\textwidth}{!}{
  \includegraphics{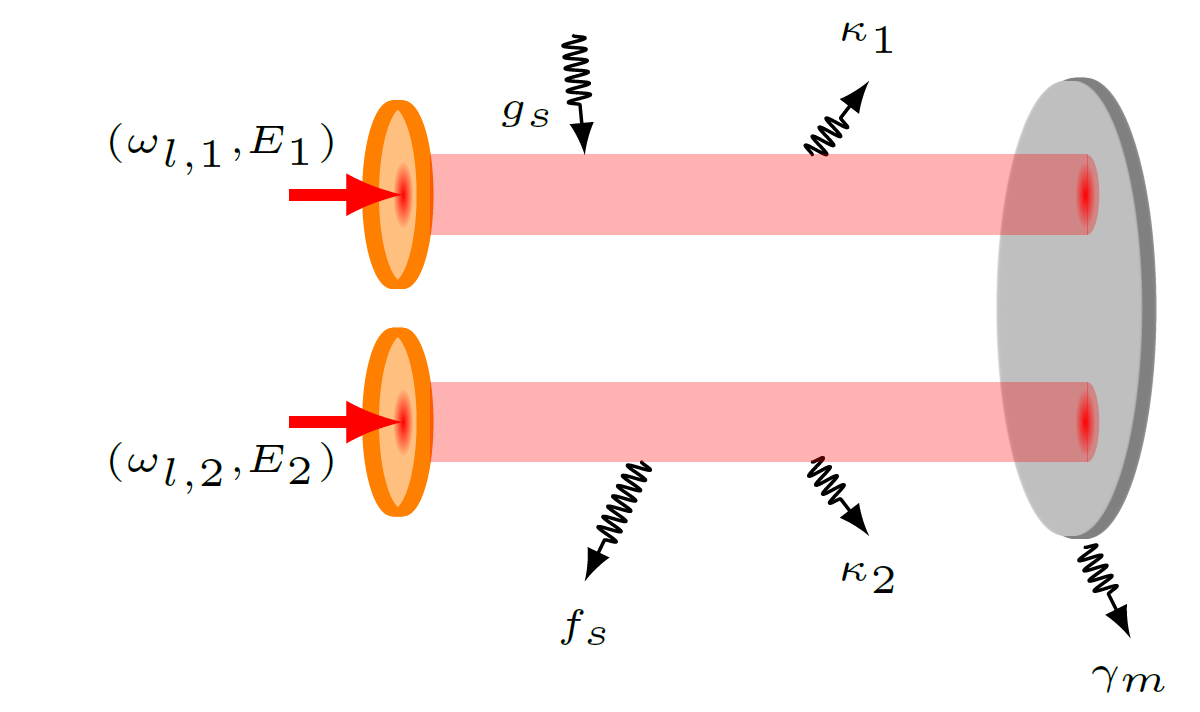}}
  \end{center}
\caption{Sketch of our benchmark model. Three mode optomechanical system made of two electromagnetic fields optically coupled, which are driving a mechanical resonator. The optical coupling is captured through the photon hopping $J$ that is phase modulated ($\theta$) under the synthetic magnetism. The optical and the mechanical dissipations are $\kappa_j$ and $\gamma_m$, respectively.}
\label{fig:Fig1}
\end{figure}

Our benchmark system is made of two electromagnetic fields which are driving a common mechanical resonator. The two electromagnetic fields are optically coupled via a photon hopping rate $J$, which is modulated through a phase $\theta$. Each  driving field forms an optical cavity with the mechanical resonator, and each cavity hosts a saturable gain or loss which can either strengthen or weakness the cavity decay rate as depicted in \autoref{fig:Fig1}.  The phase modulation of the optical coupling refers to a synthetic magnetism for photons that has intensively investigated recently both theoretically and experimentally \cite{Schmidt2015,Fang2017,Brendel2017,Mathew2020,Chen2021,Slim2025}. For intance, such a phase modulation can be seen as a phase difference of the involved driving fields. The dynamical state of our system can be captured by the following Hamiltonian ($\hbar=1$),

\begin{align}
H&=\sum_{j=1,2} \omega_{c,j} a_j^\dagger a_j + \omega_m b^\dagger b +\sum_{j=1,2} g_j(b^\dagger + b)a_j^\dagger a_j \nonumber \\ &+ J(a_1^\dagger a_2 e^{i\theta} + a_1 a_2^\dagger e^{-i\theta}) + \sum_{j=1,2} E_j(a_j^\dagger e^{-i\omega_{l,j}t} + a_j e^{i\omega_{l,j}t}) \nonumber \\ & + ig_s a_1^\dagger a_1 - i f_s a_2^\dagger a_2,
\end{align}
where $a_{j=1,2}$ and $b$ are the bosonic modes for the $j^{th}$ cavity field and the mechanical resonator, respectively. The $j^{th}$ cavity and the mechanical frequencies are captured by  $\omega_{c,j}$ and $\omega_m$, while $g_j$ stands for the optomechanical coupling between each optical mode and the mechanical mode. Each driving field has an amplitude $E_j$ and a frequency $\omega_{l,j}$. The saturable gain (loss) in the cavity is defined as, $g_s=g_0/(1+|\alpha_1|^2)$ ($f_s=f_0/(1+|\alpha_2|^2)$), where $|\alpha_j|$ is the $j^{th}$ steady state intracavity field.

After a rotating wave approximation in the frame of the driving frequencies, the Quantum Langevin Equations (QLEs) can be derived as, 

\begin{align}\label{eq:qle}
\small{\begin{cases} 
\dot{a}_1&=-\left(i\left(\Delta_{c,1} +g_1(b^\dagger+b)\right) - g \right)a_1-iJ a_2 e^{i\theta} -iE_1+\sqrt{2\kappa_1}a_1^{in}, \\
\dot{a}_2&=-\left(i\left(\Delta_{c,2} +g_2(b^\dagger+b)\right) + f \right)a_2-iJ a_1 e^{-i\theta} -iE_2 +\sqrt{2\kappa_2}a_2^{in}, \\
\dot{b}&=-\left(i\omega_m + \gamma_m \right)b - i\sum_{j=1,2} g_j a_j^\dagger a_j +\sqrt{2\gamma_m}b^{in}, 
\end{cases}}
\end{align}
where we have taken into account the cavity' decay rates $\kappa_j$  and the mechanical dissipation $\gamma_m$. In \autoref{eq:qle}, we have defined $g=g_s - \kappa_1$ and $f=f_s + \kappa_2$ as the net modal gain and loss in the first and second mode, respectively. Moreover, the detunings $\Delta_{c,j}=\omega_{c,j}-\omega_{l,j}$ have been included as well.

In order to capture quantum features of our system, we linearize the above nonlinear QLEs by following the usual procedure, where any operator ($\mathcal{O}\equiv a_j,b$) is split into its mean value with some amount of fluctuations, i.e., $a_j=\alpha_j + \delta a_j$ and $b=\beta + \delta b$. By plugging back these expressions in \autoref{eq:qle}, we get the mean value equations, 

\begin{align}\label{eq:mean}
\begin{cases}
\dot{\alpha}_1&=-\left(i{\Delta}_1 - g \right)\alpha_1-iJ e^{i\theta} \alpha_2 - iE_1, \\
\dot{\alpha}_2&=-\left(i{\Delta}_2 + f \right)\alpha_2-iJ e^{-i\theta} \alpha_1 - iE_2, \\
\dot{\beta}&=-\left(i\omega_m + \gamma_m \right)\beta - i\sum_{j=1,2} g_j |\alpha_j|^2,
\end{cases}
\end{align}
and the equations capturing the fluctuations,

\begin{align}\label{eq:fluct}
\small{\begin{cases}
\delta \dot{a}_1&=-\left(i{\Delta}_1 - g \right)\delta a_1 - iG_1(\delta b^\dagger + \delta b)-iJ e^{i\theta} \delta a_2 +\sqrt{2\kappa_1}\delta a_1^{in}, \\
\delta \dot{a}_2&=-\left(i{\Delta}_2 + f \right)\delta {a}_2 - iG_2(\delta b^\dagger + \delta b) -iJ e^{-i\theta} \delta a_1 +\sqrt{2\kappa_2}\delta a_2^{in}, \\
\delta \dot{b}&=-\left(i\omega_m + \gamma_m \right)\delta {b} - i\sum_{j=1,2} (G_j \delta a_j^\dagger + G_j^* \delta a_j)+\sqrt{2\gamma_m}\delta b^{in},
\end{cases}} 
\end{align} 
where we have defined the effective detuning ${\Delta}_j=\Delta_{c,j} +g_j(\beta^\ast +\beta)$ and the effective coupling $G_{j}=g_j \alpha_j$. Moreover, the noise operators $\delta\mathcal{O}^{in}$ have been introduced and we have assumed $g_s\sim g_0$ and $f_s \sim f_0$ for the linear regime in weak driving case \cite{Hassan2015}, where the nonlinear saturation mechanisms can be ignored. The noise operators are zero-mean valued and are characterized by the following auto-correlation functions,
\begin{eqnarray}\label{eq:noise}
\langle \delta a_j^{in}(t)\delta a_j^{in\dagger}(t') \rangle =&\delta(t-t'), \nonumber \\ 
\langle \delta a_j^{in\dagger}(t)\delta a_p^{in}(t') \rangle =& 0, \nonumber \\
\langle \delta b^{in}(t)\delta b^{in\dagger}(t') \rangle  =& (n_{th}+1)\delta(t-t'), \nonumber \\
\langle \delta b^{in\dagger}(t)\delta b^{in}(t') \rangle =& n_{th}\delta(t-t'),\nonumber 
\end{eqnarray}
where $n_{th}=[\rm exp(\frac{\hbar \omega_m}{k_bT})-1]^{-1}$ stands for the thermal phonon occupation of the mechanical resonator, with $\rm k_b$ the Boltzmann constant and $\rm T$ the bath temperature.

For each operator $\delta\mathcal{O}$, we associate the corresponding position and momentum quadrature operators defined as: $\delta {x}_{j,m}=\frac{\delta \mathcal{O} + \delta \mathcal{O}^\dagger}{\sqrt{2}}$, $\delta {p}_{j,m}=i\frac{\delta \mathcal{O}^\dagger - \delta \mathcal{O}}{\sqrt{2}}$. Similarly, the corresponding quadratures for noise operators are $\delta {x}_{j,m}^{in}=\frac{\delta \mathcal{O}^{in} + \delta \mathcal{O}^{\dagger in}}{\sqrt{2}}$ and $\delta {p}_{j,m}^{in}=i\frac{\delta \mathcal{O}^{\dagger in} - \delta \mathcal{O}^{in}}{\sqrt{2}}$. The using of these quadratures in the above fluctuation equations (\autoref{eq:fluct}) leads to the following compact form of the quadrature equations, 
\begin{equation}\label{eq:comp}
\delta\dot{u}=\rm{M}\delta u+ \rm{N} \delta u^{in},
\end{equation}
with the column vector $\delta u=(\delta {x}_1, \delta {p}_1,\delta {x}_2, \delta {p}_2,\delta {x}_m, \delta {p}_m)^{\top}$, and its related noise vector $\delta u^{in}=(\delta {x}_1^{in}, \delta {p}_1^{in},\delta {x}_2^{in}, \delta {p}_2^{in},\delta {x}_m^{in}, \delta {p}_m^{in})^{\top}$, where the symbol ($^{\top}$) refers to a transposition operation. In \autoref{eq:comp}, the noise matrix is defined as, $\rm{N}=Diag[\sqrt{2\kappa_1}, \sqrt{2\kappa_1}, \sqrt{2\kappa_2},\sqrt{2\kappa_2},\sqrt{2\gamma_m},\sqrt{2\gamma_m}]$, and the drift matrix $\rm{M}$ reads,


\begin{equation}\label{eq:corr}
M=\begin{pmatrix}
g&\Delta_1&J\sin\theta&J\cos\theta&0&0 \\
-\Delta_1&g&-J\cos\theta&J\sin\theta&-2G_{1}&0 \\
-J\sin\theta&J\cos\theta&-f&\Delta_2&0&0 \\
-J\cos\theta&-J\sin\theta&-\Delta_2&-f&-2G_{2}&0 \\
0&0&0&0&-\gamma_m&\omega_m \\
-2G_{1}&0&-2G_{2}&0&-\omega_m&-\gamma_m
\end{pmatrix},
\end{equation}
where we  have assumed $G_j$ as real.

\begin{figure}[tbh]
\begin{center}
  \resizebox{0.45\textwidth}{!}{
  \includegraphics{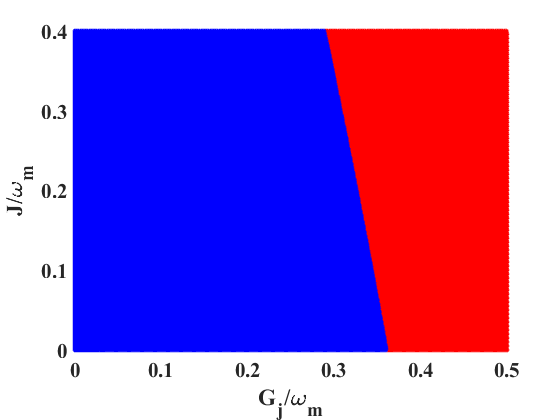}}
  \end{center}
\caption{Stability diagram versus the optical coupling $J$ and the effective coupling $G_j$. The dark (blue) area is stable, while the red (light) area is unstable. The other parameters are $\kappa_j=0.2\omega_m$, $\gamma_m=10^{-5}\omega_m$, $\Delta_j=\omega_m$, $g_s=0$, and $f_s=0$.}
\label{fig:Fig2}
\end{figure}

To investigate both tripartite entanglement and quantum coherence, we compute the covariance matrix $\rm V$ whose elements $\rm{V}_{i,j}$ are defined as $\rm{V}_{i,j}=\frac{u_iu_j + u_ju_i}{2}$, which satisfy the motional equation,
\begin{equation}\label{eq:lyapp}
    \rm{\dot{V}=MV+VM^{\top}+D},
\end{equation}
with the diffusion matrix $\rm{D}=Diag[\kappa_1, \kappa_1, \kappa_2, \kappa_2, \gamma_m(2n_{th}+1), \gamma_m(2n_{th}+1)]$. As we are interested to stationary entanglement, the matrix $\rm{M}$ must be stable, i.e., all its eigenvalues should have negative real parts as required by the Routh-Huritz stability criterion \cite{DeJesus}. Under such stability condition, the matrix elements of $\rm{V}$ should reach their steady-state and \autoref{eq:lyapp} reduces to the following  Lyapunov equation,
\begin{equation}\label{eq:lyap}
    \rm{MV+VM^{\top}=-D}.
\end{equation}
The covariance matrix is extracted from \autoref{eq:lyap}, and it has the general form, 
\begin{equation}\label{eq:cov}
\rm{V=
\begin{pmatrix} 
V_{\alpha_1}&V_{\alpha_1,\alpha_2}&V_{\alpha_1,\beta} \\
V_{\alpha_1,\alpha_2}^{\top}&V_{\alpha_2}&V_{\alpha_2,\beta} \\ 
V_{\alpha_1,\beta}^{\top}&V_{\alpha_2,\beta}^{\top}&V_{\beta}
\end{pmatrix}},
\end{equation}
where $\rm V_i$ and  $\rm V_{ij}$ are $2\times2$ block of  matrices (with $i,j \equiv \alpha_1,\alpha_2,\beta$). The elements of the covariance matrix can be analytically evaluated, but this leads to a tedious task and cumbersome expressions. For seek of simplicity, these expressions will be computed numerically instead.

\section{Quantification of tripartite entanglement and quantum coherence}\label{sec:metr}
This section provides metrics that will be used to quantify tripartite entanglement and quantum coherence in our investigation. For the tripartite entanglement, we will use the minimal residual contangle $R_{min}$ defined as \cite{Adesso2006,Adesso2007},
\begin{equation}\label{eq:resid}
    R_{min}={\rm min}[E_\tau^{r|st}-E_\tau^{r|s}-E_\tau^{r|t}],
\end{equation}
where $(r,s,t)\in\{a_1,a_2,b\}$ are the possible permutations to perform. The quantity $E_\tau^{m|n}$ stands for the contangle between subsystem $m$ (only one mode) and subsystem $n$ (containing one or two modes), and is defined as the logarithmic negativity squared,
\begin{equation}\label{eq:contang}
    E_\tau^{m|n}=(E_N)^2 ,
\end{equation}
with 
\begin{equation}\label{eq:neglog}
    E_N={\rm max} [0,-\ln(2\nu)] .
\end{equation}
When $n$ is reduced to only one mode, the minimum
symplectic eigenvalue $\nu$ of the partial transpose of a reduced $4\times4$ covariance matrix $V_{m|n}$ is given by $\nu=\frac{1}{\sqrt{2}}\sqrt{\sum(V_{m|n})-\sqrt{\sum(V_{m|n})^2-\det{V}_{m|n}} }$ and $\sum(V_{m|n})=\det{V}_m + \det{V}_n-2\det{V}_{m,n}$. $V_{m|n}$ has the form,
\begin{equation}
    V_{m|n}=\begin{pmatrix}
        V_m&V_{m,n} \\
        V_{m,n}^\top&V_n
    \end{pmatrix}.
\end{equation}
For $n$ containing two modes, the expression of $\nu$ becomes $\nu={\rm min} [{\rm eig}|i\Omega_3V_{m|n}|]$, with $\Omega_3=\bigoplus_{k=1}^3 (i\sigma_y)$ and $\sigma_y$ is the $y-$Pauli matrix. In this case, $V_{m|n}$ is the partial transpose of the covariance matrix $V$, following the relation $V_{m|n}=P_{m|n}VP_{m|n}$. The matrices $P_{m|n}$ corresponding to the quantities $E_\tau^{r|st}$, $E_\tau^{s|rt}$ and $E_\tau^{t|rs}$ are  $P_{r|st}={\rm diag}[1,-1,1,1,1,1]$, $P_{s|rt}={\rm diag}[1,1,1,-1,1,1]$ and $P_{t|rs}={\rm diag}[1,1,1,1,1,-1]$, respectively. Therefore, $R_{min}>0$ is the necessary condition for a tripartite quantum entanglement.

In order to quantify the quantum coherence, we need to define the first and second order moments in our system. For $\rm N$ bosonic modes, the first moment vector reads $\overrightarrow{D}=[\overrightarrow{d_1},\cdots,\overrightarrow{d_i},\cdots, \overrightarrow{d_N}]$, where the first moment of the $i^{th}$ bosonic mode is $\overrightarrow{d_i}$. The second order moments are extracted from the covariance matrix $\rm V$, and the covariance matrix related to the $i^{th}$ bosonic mode is $\rm V_i$. For instance, $\overrightarrow{d_{\beta}}=[\langle x_{\beta} \rangle,\langle p_{\beta} \rangle]$, where $\langle \bullet \rangle$ refers to a mean value and  $x_{\beta}=(\beta_s + \beta_s^{\ast})/\sqrt{2}$, $p_{\beta}=(\beta_s^{\ast} - \beta_s)/\sqrt{2}$ with $\beta_s$ being the steady state value of the mechanical mode extracted from \autoref{eq:mean}, i.e., for $\dot \beta =0$. Similarly, the covariance matrix of the mechanical mode is captured by $\rm V_{\beta}$ (from \autoref{eq:cov}). To further easy our discussion on a subsequent quantum coherence, we define the following submatrices,

\begin{equation}\label{eq:a1a2}
\rm{V_{\alpha_1,\alpha_2}}=
\begin{pmatrix} 
V_{\alpha_1}&V_{\alpha_1,\alpha_2} \\
V_{\alpha_1,\alpha_2}^{\top}&V_{\alpha_2}
\end{pmatrix},
\end{equation}

\begin{equation}\label{eq:a1b}
\rm{V_{\alpha_1,\beta}}=
\begin{pmatrix} 
V_{\alpha_1}&V_{\alpha_1,\beta} \\
V_{\alpha_1,\beta}^{\top}&V_{\beta}
\end{pmatrix},
\end{equation}

and 

\begin{equation}\label{eq:a2b}
\rm{V_{\alpha_2,\beta}}=
\begin{pmatrix} 
V_{\alpha_2}&V_{\alpha_2,\beta} \\
V_{\alpha_2,\beta}^{\top}&V_{\beta}
\end{pmatrix},
\end{equation}
which capture interaction between two bosonic modes. Therefore, the quantum coherence for $\rm N$ modes is quantified as \cite{Zheng2016,Peng2024},

\begin{equation}\label{eq:coh}
 C(\rm V)=\sum_{i=1}^{\rm N} F(2n_i +1) - F(\eta_i),
\end{equation}
with 
\begin{equation}
 n_i= [\rm Tr( V_i)+d_{i,x}^2+d_{i,y}^2-2]/4,
\end{equation}
and 
\begin{equation}
 F(x)=\frac{x+1}{2}\ln{\left( \frac{x+1}{2}\right)} - \frac{x-1}{2}\ln{\left( \frac{x-1}{2}\right)},
\end{equation}
where $\rm Tr$ performs the trace operation, $F(x)$ evaluates the entropy of a Gaussian system, $d_{i,x}$, and $d_{i,y}$ capture the amplitude and the phase of the subsequent first moment, respectively. The quantity $\eta$ ($\eta_i\in \eta $) are the symplectic spectrum of the covariance matrix $\rm V$, which can be extracted from the eigenspectrum of $|\rm i\Omega V|$, where  $\Omega$ is a $2\rm N \times 2\rm N$ symplectic matrix constructed as, 
\begin{equation}
\Omega=\sum_{i=1}^{\rm N}\oplus \omega_i, \hspace{1em} 
\rm{\omega_i}\equiv
\begin{bmatrix} 
0&1 \\
-1&0
\end{bmatrix}.
\end{equation}

While \autoref{eq:coh} is suitable for quantifying quantum coherence of any Gaussian state, the one-mode quantum coherence is simply quantified as,

\begin{equation}\label{eq:coh1}
 C_{i=\alpha_1,\alpha_2,\beta}(\rm V_i)= F(2n_i +1) - F(\eta_i),
\end{equation}
where $\eta_i=\sqrt{\det{(\rm V_i)}}$. Moreover, the two-mode Gaussian quantum coherence is evaluated through,

\begin{equation}\label{eq:coh2}
 C_{ij}(\rm V_{ij})=\sum_{k=i,j} F(2n_k +1) - \sum_{\mu=\pm}F(\eta_{ij,\mu}),
\end{equation}
where $\eta_{ij,\mu}$ ($\mu=\pm$) is the symplectic eigenvalue of any submatrix $\rm V_{ij}$ as aforementioned (see \autoref{eq:a1a2}, \autoref{eq:a1b}, and \autoref{eq:a2b}), and yields $\eta_{ij,\pm}=\frac{1}{\sqrt{2}} \left[\rm  \Gamma_{ij}\pm \sqrt{\rm  \Gamma_{ij}^2-4\det{(\rm V_{ij})}}\right]^{1/2}$, with $\rm \Gamma_{ij}=\det{(\rm V_i)}+\det{(\rm V_j)}+2\det{(\rm V_{ij})}$. From now on, we will use $C_{\alpha_1,\alpha_2}=C(V_{\alpha_1,\alpha_2})$, $C_{\alpha_1,\beta}=C(V_{\alpha_1,\beta})$, $C_{\alpha_2,\beta}=C(V_{\alpha_2,\beta})$, and $C_t=C(V)$ to represent the quantum coherence between the two optical modes, the first optical and the mechanical modes, the second optical and the mechanical modes, and the full three modes, respectively.

\begin{figure*}[tbh]
\begin{center}
  \resizebox{0.45\textwidth}{!}{
  \includegraphics{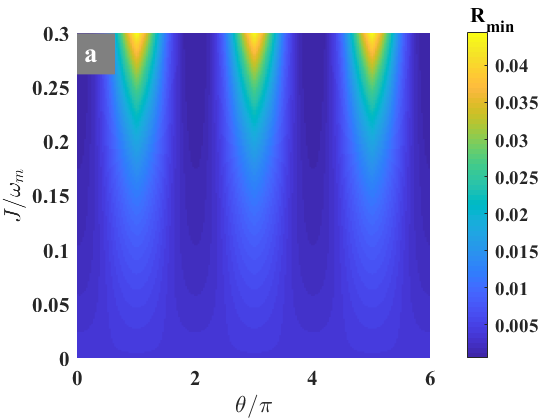}}
   \resizebox{0.45\textwidth}{!}{
  \includegraphics{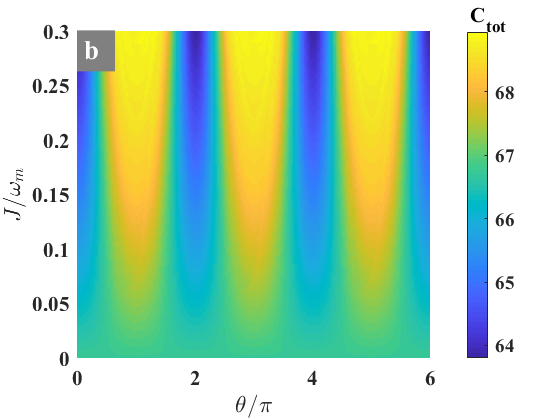}}
  \resizebox{0.45\textwidth}{!}{
  \includegraphics{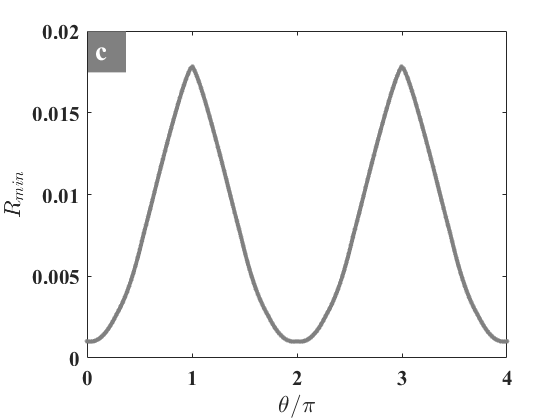}}
   \resizebox{0.45\textwidth}{!}{
  \includegraphics{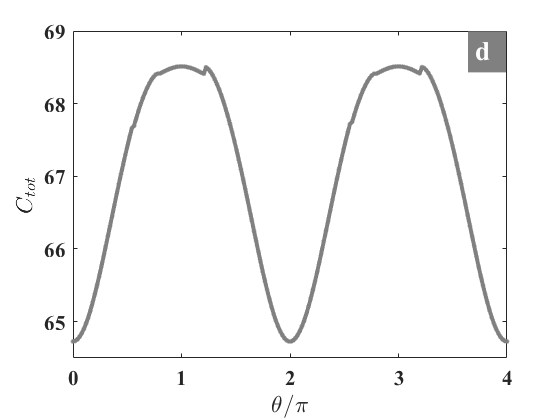}}
  \end{center}
\caption{Contour plot of tripartite entanglement (a) and quantum coherence (b) versus the optical coupling $J$ and its phase modulation $\theta$. (c) and (d) show $2D$-representations extracted from (a) and (b), respectively for $J=0.2\omega_m$. The other parameters are $\kappa_j=0.2\omega_m$, $\gamma_m=10^{-5}\omega_m$, $G_j=0.15\omega_m$, $n_{th}=100$, $\Delta_j=\omega_m$, $g_s=0$, and $f_s=0$.}
\label{fig:Fig3}
\end{figure*}

\section{effect of the optical coupling on both entanglement and quantum coherence}\label{sec:coupl}
We first investigate the effect of the optical coupling, captured by the photon hopping rate $J$ and its modulation phase $\theta$, on both the tripartite entanglement and the quantum coherence. The idea is to figure out the common features exhibited by entanglement and quantum coherence which are induced via the optical coupling. For this purpose, we will use the following experimentally feasible parameters \cite{Fang2017,Mathew2020,Nori2022,Slim2025}, $\omega_m=10\rm{MHz}$, $\kappa_j=0.2\omega_m$, $\gamma_m=10^{-5}\omega_m$, $g_j=10^{-4}\omega_m$, and we consider working at the blue-sideband $\Delta_j=\omega_m$. The other parameters will be mentionned onwards. To be meaninfull, our used parameters must fulfill stability criterion since we are interested on stationary entanglement. The stability diagram of our system is depicted on \autoref{fig:Fig2} depending on both the optical coupling $J$ and the effective optomechanical coupling $G_j$, and our used parameters have been chosen accordingly. The dark (blue) area is  stable, while the light (red) zone is unstable. This stability diagram shows that strong input driving leads to instability ($G_j\gtrsim 0.3\omega_m$), while the optical coupling can be flexibly adjusted. From now on, we will choose both $G_j$ and  $J$ around $0.2\omega_m$ for our analysis.  

To point out the effect of the modulation phase  on both the entanglement and quantum coherence, we represent on \autoref{fig:Fig3} these quantities depending on $\theta$. It can be seen that the optimal values of these quantities are reached for $\theta=n\pi$, $n$ being an odd enteger (see \autoref{fig:Fig3} (a,b)). Conversely, the minimal values of these quantities happen for $\theta=m\pi$, $m$ being an even enteger. We also observe that the residual cotangle and the quantum coherence increase as the photon hopping rate gets stronger. By setting $J=0.2\omega_m$, we have extracted the entanglement (see \autoref{fig:Fig3}c) from \autoref{fig:Fig3}a, and the quantum coherence (see \autoref{fig:Fig3}d) from \autoref{fig:Fig3}b. It can be seen that the peaks reached by the residual cotangle are sharper compared to those of the quantum coherence, which are smoother. This means that the optimal value of the entanglement is reached for a fine tuning of the phase $\theta$, while getting optimal value of the quantum coherence does not require a rigorous control over the phase. Moreover,  \autoref{fig:Fig3}d shows that the coherence forms a sort of plateau around $\theta=n\pi$, meaning that optimal quantum coherence can be reached for other values of the phase around  $\theta=n\pi$. From the similarities pointed out between the tripartite entanglement and the quantum coherence, it is worth to remark that residual cotangle is displayed over a narrow range of  $\theta$ comapred to the quantum coherence.  In the following, we will use $\theta=\pi$ to ensure staying near the optimal engineered values of our residual cotangle and quantum coherence.  

\begin{figure}[tbh]
\begin{center}
  \resizebox{0.45\textwidth}{!}{
  \includegraphics{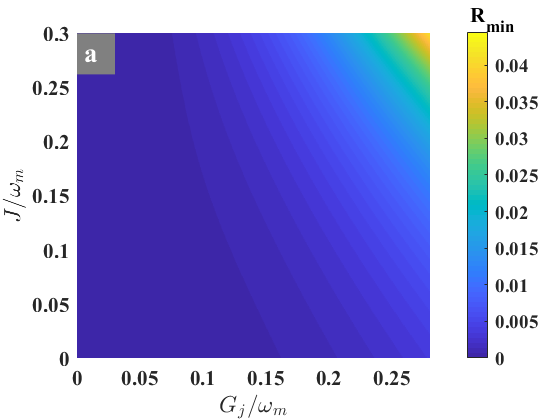}}
  \resizebox{0.45\textwidth}{!}{
  \includegraphics{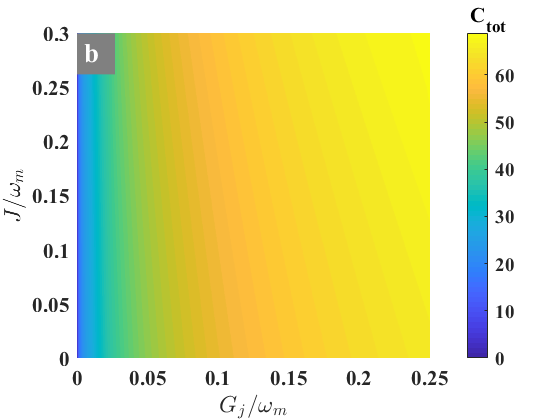}}
  \end{center}
\caption{(a) Contour plot of tripartite entanglement,  and (b) contour plot of the quantum coherence versus $J$ and $G_j$ for $\theta=\pi$. The other parameters are the same as those in \autoref{fig:Fig3}.}
\label{fig:Fig4}
\end{figure}

\autoref{fig:Fig4} displays the residual cotangle (see \autoref{fig:Fig4}a) and the quantum coherence (\autoref{fig:Fig4}b) versus the photon hopping rate $J$ and the effective optomechanical coupling $G_j$. It can be observed that both tripartite entanglement and quantum coherence increase over the coupling $G_j$. This means that these quantities are improved as the driving strengths get stronger. It is worth mentioning that the entanglement is generated from a certain threshold of the driving strengths, which correspond to an optomechanical coupling strengths around $G_j=0.1\omega_m$ (see \autoref{fig:Fig4}a). However, the threshold corresponding to a generation of quantum coherence happens very earlier compared to the one for the residual cotangle. Even though both quantities are enhanced over the driving strengths in \autoref{fig:Fig4}, it is worth to underline that the generation of  quantum coherence is less power demanding compared to the tripartite entanglement. As aforementioned regarding the phase's effect, the engineering of the coherence is more flexible with relaxed condition upon the used input powers. 

\begin{figure}[tbh]
\begin{center}
  \resizebox{0.45\textwidth}{!}{
  \includegraphics{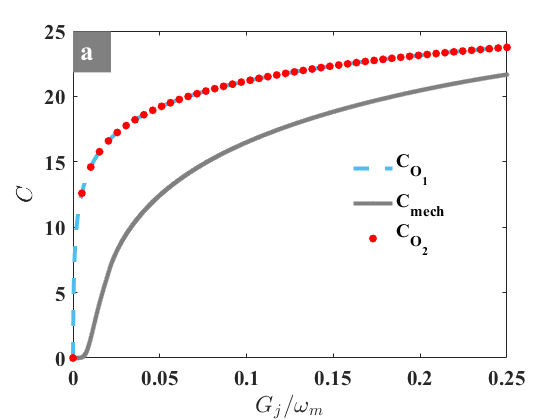}}
  \resizebox{0.45\textwidth}{!}{
  \includegraphics{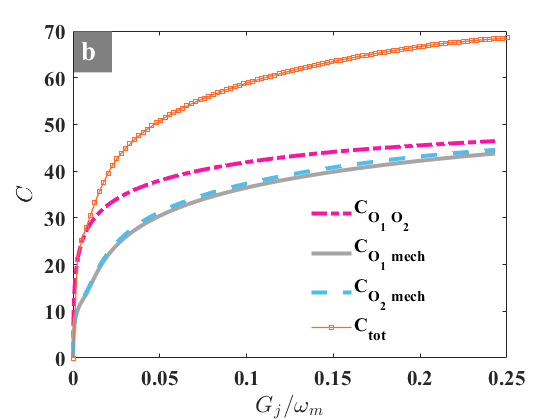}}
  \end{center}
\caption{(a) one mode, and (b) multi-modes  quantum coherences, versus the effective optomechanical coupling $G_j$ for $J=0.2\omega_m$.  The other parameters are the same as those in \autoref{fig:Fig3}.}
\label{fig:Fig5}
\end{figure}

\begin{figure*}[tbh]
\begin{center}
  \resizebox{0.45\textwidth}{!}{
  \includegraphics{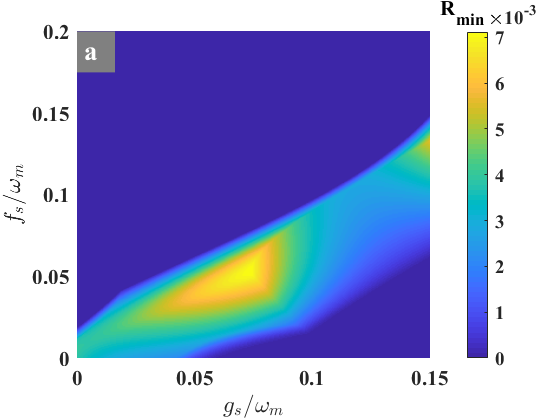}}
  \resizebox{0.45\textwidth}{!}{
  \includegraphics{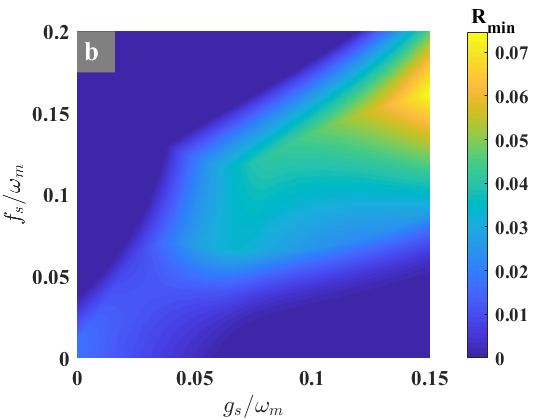}}
  \resizebox{0.45\textwidth}{!}{
  \includegraphics{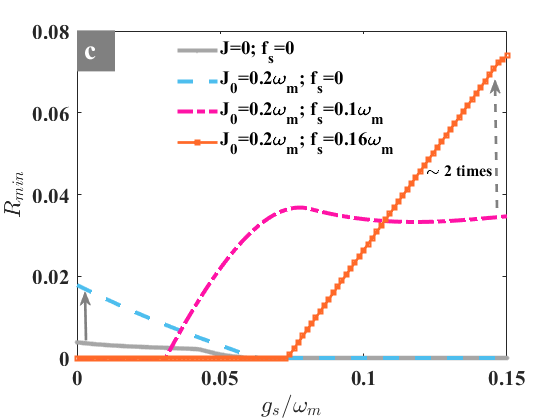}}
  \resizebox{0.45\textwidth}{!}{
  \includegraphics{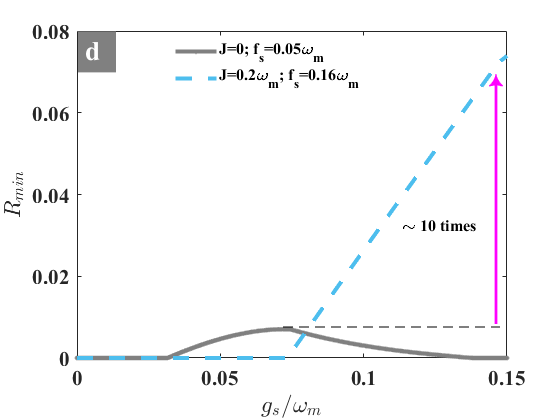}}
  \end{center}
\caption{Contour plot of tripartite entanglement, without (a) and with (b) the optical coupling $J$ versus saturable gain ($g_s$) and loss ($f_s$), for $G_j\sim 0.2 \omega_m$ and $\theta=\pi$. (c) $2D$-representation of the entanglement extracted from (a) and (b) for some specific parameters. (d) optimal entanglement extracted from (a) and (b), i.e., $J=0$, $f_s=0.05\omega_m$ (full line), and $J=0.2\omega_m$, $f_s=0.16\omega_m$ (dashed line). The other parameters are the same as those in \autoref{fig:Fig3}.}
\label{fig:Fig6}
\end{figure*}
For a seek of completeness on quantum coherence, we have investigated on one and two modes quantum coherences involved in our system. Such analysis are carried out on \autoref{fig:Fig5}. Indeed, one mode quantum coherences are displayed in \autoref{fig:Fig5}a. The full line captures the mechanical quantum coherence, while the dashed and dotted lines depict the first and the second optical mode quantum coherence, respectively. We observe that the mechanical quantum coherence is less than the two optical coherences, which are similar. Such a result is related to the fact that the two optical cavities are coupled to each other through the coupling $J$, which is generaly greater than the optomechanical $G_j$ in the linear regime as considered for our analysis. Therefore, it is reasonable to have higher optical quantum coherence, compared to the mechanical coherence. For strong coupling $G_j$, the mechanical quantum coherence may become comparable to the optical one (see \autoref{fig:Fig5}a from $G_j\gtrsim0.25\omega_m$). However, we stress that our system will get into the nonlinear regime  as depicted in \autoref{fig:Fig2}. The two modes quantum coherences involved in our system are displayed in \autoref{fig:Fig5}b versus the effective optomechanical coupling. The dashed line and the associated light full line capture the quantum coherences between the optical and mechanical modes. The dot-dashed line represent quantum coherence between the two optical modes, while the squared line shows the tripartite quantum coherence. What is observed is that the coherences involving one optical and mechanical modes are similar, while quantum coherence between optical modes is greater compared to those involving mechanical mode. Moreover, it can be seen that the amound of tripartite quantum coherence is larger compared to the lower mode quantum coherences. This comes from the definitions of different types of quantum coherences as stated above. In fact, the more the mode are involved within the quantum coherence, the more are the sum of terms involved. This is why we can observe the following hierarchy on \autoref{fig:Fig5}, i.e., the amount of the tripartite quantum coherence is larger than the amounts of two modes quantum coherences, which in turn are higher compared to the amount of the one mode quantum coherences. Other aspects to mention are the fact that quantum coherences increase upon increasing the driving strengths (or $G_j$), and the optical coupling $J$ does not significantly affect the quantum coherence (checked but not shown here).     

\begin{figure*}[tbh]
\begin{center}
  \resizebox{0.45\textwidth}{!}{
  \includegraphics{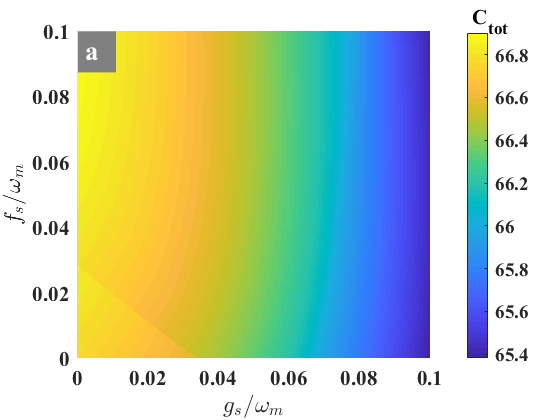}}
  \resizebox{0.45\textwidth}{!}{
  \includegraphics{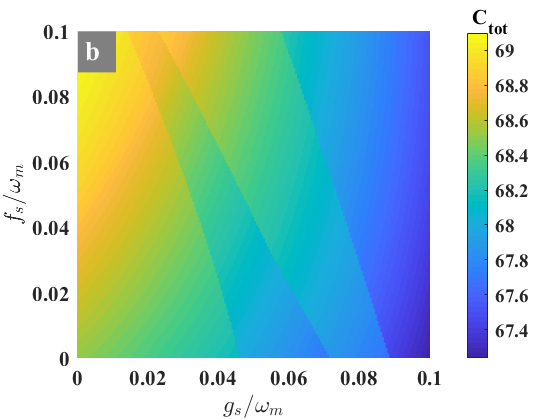}}
  \resizebox{0.45\textwidth}{!}{
  \includegraphics{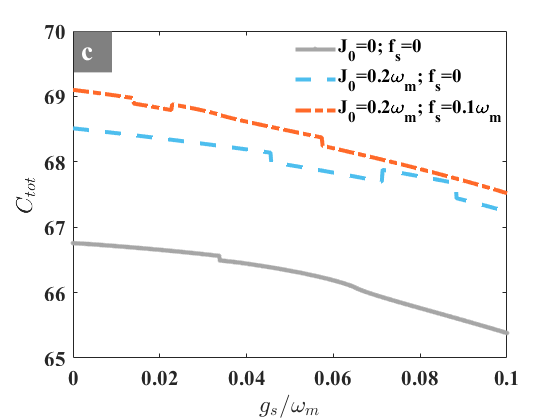}}
  \resizebox{0.45\textwidth}{!}{
  \includegraphics{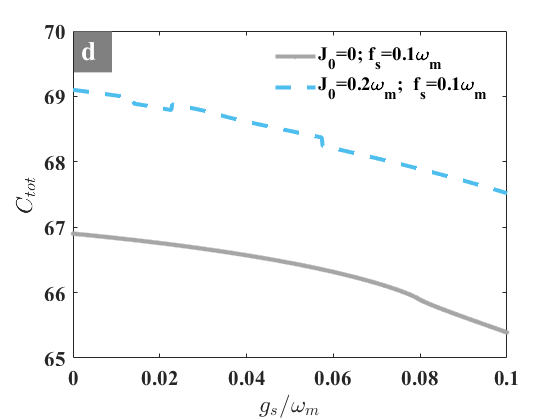}}
  \end{center}
\caption{Contour plot of three modes quantum coherence without (a) and with (b) the optical coupling $J$ versus saturable gain ($g_s$) and loss ($f_s$), for $G_j\sim 0.2 \omega_m$ and $\theta=\pi$. (c) $2D$-representation of the quantum coherence extracted from (a) and (b) for some specific parameters. (d) optimal quantum coherence extracted from (a) and (b), i.e., $J=0$, $f_s=0.1\omega_m$ (full line), and $J=0.2\omega_m$, $f_s=0.1\omega_m$ (dashed line). The other parameters are the same as those in \autoref{fig:Fig3}. The other parameters are the same as those in \autoref{fig:Fig3}.}
\label{fig:Fig7}
\end{figure*}

\section{Saturable gain effect on both entanglement and quantum coherence}\label{sec:sat}
In this section, we are going to take into account the saturable gain/loss effect in our analysis in order to further control both the generation of the tripartite entanglement and the quantum coherence. As we have limited our investigation within the linear regime, only the linear part of the saturation mechanisms have been considered as aforementioned. The full saturable gain/loss effect will be accounted in our future work, i.e., beyond the linearized regime. Therefore, we represent on \autoref{fig:Fig6}(a,b) the minimal residual contangle versus the saturable gain and loss. \autoref{fig:Fig6}a is plotted for $J=0$, while \autoref{fig:Fig6}b corresponds to an optical coupling of $J=0.2\omega_m$. It can be seen that a modest degree of tripartite entanglement is generated even without the introduction of saturation effects (see \autoref{fig:Fig6}a at $g_s=f_s=0$). As the saturation effects are incorporated, one observes an enhancement of tripartite entanglement, particularly when the saturable gain tends to compensate the saturable loss ($g_s\approx f_s$). Therefore, a good matching between saturable gain and losses leads to an efficient control of tripartite entanglement engineering in our proposal. Under the joint effect of both optical coupling and saturable effect, the resulted entanglement is further enhanced as depicted in  \autoref{fig:Fig6}b. Once more, it can be seen that the maximum degree of tripartite entanglement occurs when the saturable gain and loss are approximately of the same order of magnitude. Another feature to mention is the significant entanglement enhancement in our scheme under saturable gain/loss control. Indeed, by comparing the optimal entanglement obtained in \autoref{fig:Fig3}a and \autoref{fig:Fig4}a (for larger $J$ and $G_j$) to what is reached in \autoref{fig:Fig6}b (for less strength of $J$ and $G_j$), it results that the entanglement has been improved almost twice under saturable gain/loss control. More interestingly, the overall enhancement efficiency of the entanglement in our proposal is up to one order of magnitude compared to the case without saturation effects nor optical coupling. This can be seen by comparing the colorbars of \autoref{fig:Fig6}a and \autoref{fig:Fig6}b.

In order to effectively quantify the effect of the saturable effect and those from the optical coupling, we have plotted \autoref{fig:Fig6}(c,d), which are extracted from \autoref{fig:Fig6}(a,b) for specific parameters. On \autoref{fig:Fig6}c, we have displayed different curves for different set of parameters. Indeed, the full line corresponds to $J=0$ and $f_s=0$, and the dashed curve is for $J=0.2\omega_m$ and $f_s=0$. These two curves point out the effect of the optical coupling $J$, which has enhanced the tripartite entanglement (up to $\sim 5$ times) in the absence of any saturation effect ($f_s=g_s=0$, see the full arrow line). When the saturation effects are accounted, the dash-dotted line ($J=0.2\omega_m$ and $f_s=0.1\omega_m$) and the squared line ($J=0.2\omega_m$ and $f_s=0.16\omega_m$) reveal how entanglement can be efficiently controlled and enhanced (see the dashed arrow). It results that saturable gain/loss can effectively enhance tripartite entanglement.  Moreover, \autoref{fig:Fig6}d depicts the optimal residual cotangle extracted from \autoref{fig:Fig6}a and \autoref{fig:Fig6}b. It can be seen that a joint effect of both saturation effect and optical coupling can be used to enhance the generated tripartite entanglement up to one order of magnitude as observed from the colorbars of \autoref{fig:Fig6} (a,b). Furthermore, a comparison between the light full line  and the squared line in \autoref{fig:Fig6}c reveals that the entanglement can be enhanced up to one order of magnitude in our proposal.   

\begin{figure}[tbh]
\begin{center}
  \resizebox{0.45\textwidth}{!}{
  \includegraphics{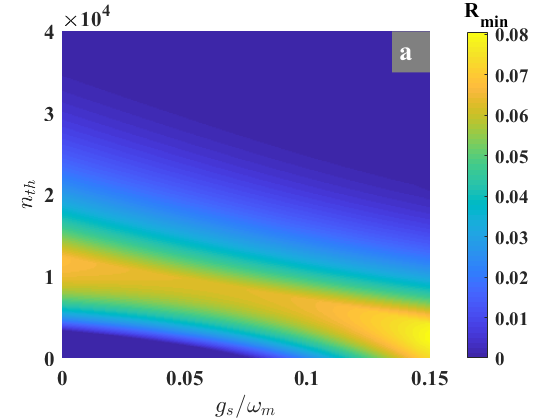}}
   \resizebox{0.45\textwidth}{!}{
  \includegraphics{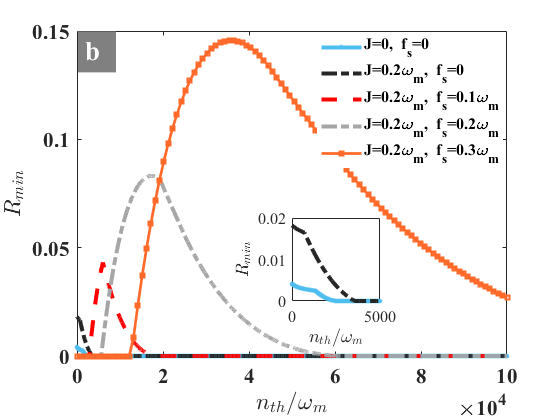}}
  \end{center}
\caption{(a) Contour plot of tripartite entanglement vs the thermal phonon number $n_{th}$ and the saturable gain $g_s$ for $f_s=0.16\omega_m$ and $J=0.2\omega_m$. (b) $2-D$ representation of the entanglement extracted from (a) versus thermal phonon number at $g_s=0$ and for specific values of $f_s$, i.e., $f_s=0$ (see  zoomed erea), $f_s=0.1\omega_m$ (dashed line), $f_s=0.2\omega_m$ (dash-dotted line), and $f_s=0.3\omega_m$ (squared line). The rest of the parameters are the same as those in \autoref{fig:Fig3}.} 
\label{fig:Fig8}
\end{figure}

To figure out the effects of both optical coupling and the saturable gain/loss on the quantum coherence, we proceed as for the entanglement done above. This is shown in \autoref{fig:Fig7}, where the three modes quantum coherence is displayed versus the saturable parameters $g_s$ and $f_s$. \autoref{fig:Fig7}a shows the quantum coherence for $J=0$, and it can be seen that saturable effects can be used to control coherence in our system. For instance, the intensity of quantum coherence decreases as the saturable gain effect increases. Conversely, the coherence is slightly improved over the increase of $f_s$. By including the optical coupling $J$, we observe an enhancement of the quantum coherence that is slightly distorded over the saturable parameters. To further figure out these effects, we plotted \autoref{fig:Fig7}c and  \autoref{fig:Fig7}d. The full line in \autoref{fig:Fig7}c corresponds to $J=0$ and $f_s=0$, while the dashed line is for $J=0.2\omega_m$ and $f_s=0$. It can be seen that the quantum coherence is improved as the coupling $J$ is taken into account. One can equaly observe a slightly enhancement of the coherence depending on the saturable loss $f_s$ (see the dash-dotted line). In  \autoref{fig:Fig7}d, we represented the optimal quantum coherence extracted from \autoref{fig:Fig7}a (full line) and from \autoref{fig:Fig7}b (dashed line). This figure captures the joint effects of both coupling $J$ and saturable parameters on the quantum coherence. In general, it is found that the quantum coherence is weakly improved under the considered parameters ($J$, $f_s$, and $g_s$), while the  tripartite entanglement has been significantly enhanced. However, either the quantum coherence or the entanglement are both controlled by the saturable effects and the optical coupling. This control scheme based on the saturable gain/loss provide a new tool to generate suitable quantum resources which may be useful for quantum technologies.  

\begin{figure}[tbh]
\begin{center}
  \resizebox{0.45\textwidth}{!}{
  \includegraphics{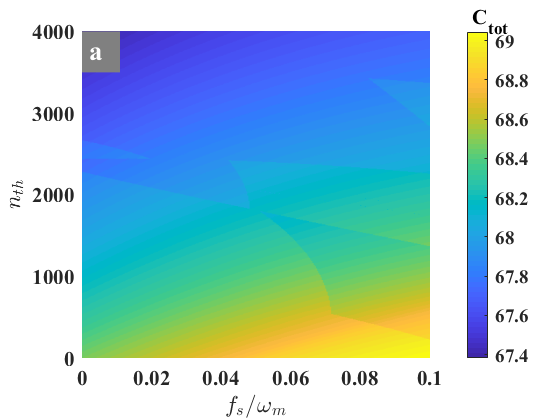}}
   \resizebox{0.45\textwidth}{!}{
  \includegraphics{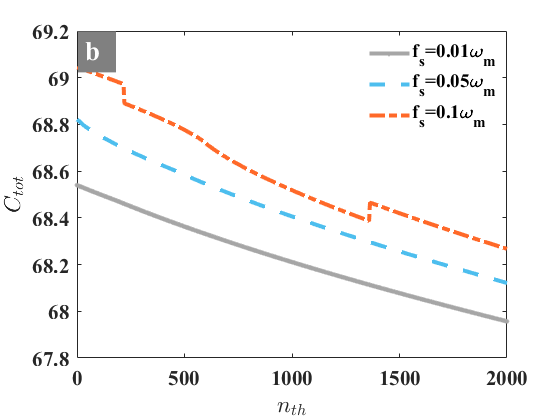}}
  \end{center}
\caption{(a) Contour plot of quantum coherence vs the thermal phonon number $n_{th}$ and the saturable loss $f_s$ for $g_s=0.01\omega_m$ and $J=0.2\omega_m$. (b) $2-D$ representation of the quantum coherence extracted from (a) versus thermal phonon number for specific values of $f_s$, i.e., $f_s=0.01\omega_m$ (full line), $f_s=0.05\omega_m$ (dashed line), and $f_s=0.1\omega_m$ (dash-dotted line). The rest of the parameters are the same as those in \autoref{fig:Fig3}.}
\label{fig:Fig9}
\end{figure}

Another interesting feature to point out is the robustness of the generated tripartite entanglement and quantum coherence against thermal fluctuations. This analysis reveals how strong the engineered quantum resources resist to thermal noise. For this purpose, we plotted \autoref{fig:Fig8} which illustrates the effect of thermal noise on tripartite entanglement. \autoref{fig:Fig8}a displays the tripartite entanglement versus  thermal phonon number $n_{th}$ and the saturable gain $g_s$ for a fixed value of the saturable loss at $f_s=0.16\omega_m$ (that corresponds to optimal entanglement reached in \autoref{fig:Fig6}b). It can be seen that the generated entanglement exhibits strong resilience to thermal noise in the presence of the saturation effect. Indeed, the engineered quantum correlation survives for thermal noise far beyond the thermal phonon occupation of $n_{th}=3\times10^{4}$, which is interesting for quantum applications at room temperature. However, it can be observed that this high resilience against thermal fluctuations slightly reduces as the saturation parameter $g_s$ increases. To further illustrate the impact of the saturation effect on the robustness of the entanglement against thermal fluctuations, we display on \autoref{fig:Fig8}b the residual cotangle versus thermal phonon number for certain values of $f_s$ for a fixed $g_s=0$. The full and the dash-dotted lines, which are zoomed out have been both plotted for $f_s=0$, and they correspond to $J=0$ and $J=0.2\omega_m$, respectively. These two curves figure out how the optical coupling $J$ strengthens the robustness of the entanglement against thermal fluctuations. It can be observed (in  zoomed erea) that the coupling $J$ slightly improves the resilience of the entanglement, which survives better in the thermal environment ($\sim n_{th}=4000$) than when there is no optical coupling ($J=0$). Under the saturation effect ($f_s\neq0$), i.e., $f_s=0.1\omega_m$ for the dashed line, $f_s=0.2\omega_m$ for the dash-dotted line, and $f_s=0.3\omega_m$ for the squared line, one can clearly see a significant enhancement of the entanglement robustness against thermal noise. It can be seen that, as the saturation loss $f_s$ increases, the entanglement survives longer to thermal fluctuations. Furthermore, the robustness of our generated entanglement is up to two order of magnitude greater than in the case without saturation effects (compare squared line with those in zoomed area). This reveals the crucial role of the saturable effects on the engineering of noise-resistant tripartite entanglement, which is the figure of merit of our proposal. Such a thermal management under saturation mechanisms is quite general and can be extended to similar systems. Furthermore, such scheme can be used as a tool for synthetizing further noise-tolerant quantum correlations.    

Similar to the entanglement, the robustness of the quantum coherence is displayed on \autoref{fig:Fig9}. As it is shown in \autoref{fig:Fig9}a, which depicts quantum coherence versus  $n_{th}$ and $f_s$, large coherences persist up to $n_{th}\sim 4000$. Conversely to the entanglement, it should be observed that quantum coherence increases over the saturable loss instead (see also \autoref{fig:Fig7}), rather than the saturable gain $g_s$. Therefore, quantum coherence is more resilient against thermal fluctuations as the saturable loss increases. This feature is better illustrated in \autoref{fig:Fig9}b, which displays quantum coherence versus $n_{th}$, for specific values of $f_s$ extracted from \autoref{fig:Fig9}a. In this figure, the full line corresponds to $f_s=0.01\omega_m$, the dashed line is for $f_s=0.05\omega_m$, while $f_s=0.1\omega_m$ corresponds to the dash-dotted line. As aforementioned, one can observe that the more  $f_s$ is larger, the more quantum coherence is robust against thermal noise. Once again, this reveals the key role of the saturable effect on the generation of quantum coherence and its robustness against thermal fluctuations. Such resilience of quantum coherence to high temperatures sheds light on the remarkable role of the saturable gain/loss for the generation of robust quantum correlations.    

\section{Conclusion}\label{sec:concl}
This work investigated on the control of both tripartite entanglement and quantum coherence through  saturable gain/loss effects and optical coupling. Our benchmark model is a mechanical resonator driven by two electromagnetic fields optically coupled. When the optical cavities are empty of the saturable gain or loss, we have shown an enhancement of both entanglement and quantum coherence under the effect of the photon hopping rate $J$, and its modulation phase $\theta$. In fact, the optimal values of these quantum resources happen at $\theta=n\pi$ with $n$ being an odd enteger, and these generated correlations increase as the optical coupling get stronger. When the cavities host the saturable gain or loss, both the tripartite entanglement and quantum coherence are not only enhanced, but they are controlled in order to generate a suitable quantum resource on demand depending on the tunability of the gain $g_s$ or loss $f_s$. Owing to the joint effect of $J$ and $g_s$ (or $f_s$), we were able to enhance tripartite entanglement up to one order of magnitude. For the quantum coherence, however, the enhancement was not significant as for the entanglement. We have also shown that the robustness of the generated tripartite entanglement against thermal noise has been significantly enhanced. Indeed, under a control of saturation loss $f_s$, the residual cotangle survives to thermal phonon number up to $n_{th}=10^5$, which is two order of magnitude stronger compared to the case without saturable effects. Such a high resilience of tripartite entanglement against thermal fluctuation suggests quantum applications at room temperatures. This reveals the key role of saturable gain/loss on the engineering of robust quantum resources, which is the figure of merit of our proposal. Our work sheds light on the using of nonlinear gain saturation as a tool for the generation of highly resilient quantum resources against thermal fluctuations. Such quantum correlations, which result from thermal management based on saturable effects, can be useful for quantum computation tasks, quantum information processing schemes, and similar quantum technologies.  

\section*{Acknowledgments}

P.D. acknowledges the Iso-Lomso Fellowship from the  Stellenbosch Institute for Advanced Study (STIAS),  Stellenbosch 7600, South Africa, and The Institute for Advanced Study, Wissenschaftskolleg zu Berlin, Wallotstrasse 19, 14193 Berlin, Germany. J.-X.P. is supported by National Natural Science Foundation of China (Grant No.~12504566), Natural Science Foundation of Jiangsu Province (Grant No.~BK20250947), Natural Science Foundation of the Jiangsu Higher Education Institutions (Grant No.~25KJB140013) and Natural Science Foundation of Nantong City (Grant No.~JC2024045). The authors are thankful to the Deanship of Graduate Studies and Scientific Research at University of Bisha for supporting this work through the Fast-Track Research Support Program.

\bibliography{Gain_saturation_Coherence}

\begin{thebibliography}{58}
\expandafter\ifx\csname natexlab\endcsname\relax\def\natexlab#1{#1}\fi
\expandafter\ifx\csname bibnamefont\endcsname\relax
  \def\bibnamefont#1{#1}\fi
\expandafter\ifx\csname bibfnamefont\endcsname\relax
  \def\bibfnamefont#1{#1}\fi
\expandafter\ifx\csname citenamefont\endcsname\relax
  \def\citenamefont#1{#1}\fi
\expandafter\ifx\csname url\endcsname\relax
  \def\url#1{\texttt{#1}}\fi
\expandafter\ifx\csname urlprefix\endcsname\relax\def\urlprefix{URL }\fi
\providecommand{\bibinfo}[2]{#2}
\providecommand{\eprint}[2][]{\url{#2}}

\bibitem[{\citenamefont{Wendin}(2017)}]{Wendin2017}
\bibinfo{author}{\bibfnamefont{G.}~\bibnamefont{Wendin}},
  \bibinfo{journal}{Reports on Progress in Physics}
  \textbf{\bibinfo{volume}{80}}, \bibinfo{pages}{106001}
  (\bibinfo{year}{2017}), ISSN \bibinfo{issn}{1361-6633}.

\bibitem[{\citenamefont{Flamini et~al.}(2018)\citenamefont{Flamini, Spagnolo,
  and Sciarrino}}]{Flamini2018}
\bibinfo{author}{\bibfnamefont{F.}~\bibnamefont{Flamini}},
  \bibinfo{author}{\bibfnamefont{N.}~\bibnamefont{Spagnolo}}, \bibnamefont{and}
  \bibinfo{author}{\bibfnamefont{F.}~\bibnamefont{Sciarrino}},
  \bibinfo{journal}{Reports on Progress in Physics}
  \textbf{\bibinfo{volume}{82}}, \bibinfo{pages}{016001}
  (\bibinfo{year}{2018}), ISSN \bibinfo{issn}{1361-6633}.

\bibitem[{\citenamefont{Slussarenko and Pryde}(2019)}]{Slussarenko2019}
\bibinfo{author}{\bibfnamefont{S.}~\bibnamefont{Slussarenko}} \bibnamefont{and}
  \bibinfo{author}{\bibfnamefont{G.~J.} \bibnamefont{Pryde}},
  \bibinfo{journal}{Applied Physics Reviews} \textbf{\bibinfo{volume}{6}},
  \bibinfo{pages}{041303} (\bibinfo{year}{2019}), ISSN
  \bibinfo{issn}{1931-9401}.

\bibitem[{\citenamefont{Pittaluga et~al.}(2025)\citenamefont{Pittaluga, Lo,
  Brzosko, Woodward, Scalcon, Winnel, Roger, Dynes, Owen, Juárez
  et~al.}}]{Pittaluga2025}
\bibinfo{author}{\bibfnamefont{M.}~\bibnamefont{Pittaluga}},
  \bibinfo{author}{\bibfnamefont{Y.~S.} \bibnamefont{Lo}},
  \bibinfo{author}{\bibfnamefont{A.}~\bibnamefont{Brzosko}},
  \bibinfo{author}{\bibfnamefont{R.~I.} \bibnamefont{Woodward}},
  \bibinfo{author}{\bibfnamefont{D.}~\bibnamefont{Scalcon}},
  \bibinfo{author}{\bibfnamefont{M.~S.} \bibnamefont{Winnel}},
  \bibinfo{author}{\bibfnamefont{T.}~\bibnamefont{Roger}},
  \bibinfo{author}{\bibfnamefont{J.~F.} \bibnamefont{Dynes}},
  \bibinfo{author}{\bibfnamefont{K.~A.} \bibnamefont{Owen}},
  \bibinfo{author}{\bibfnamefont{S.}~\bibnamefont{Juárez}},
  \bibnamefont{et~al.}, \bibinfo{journal}{Nature}
  \textbf{\bibinfo{volume}{640}}, \bibinfo{pages}{911} (\bibinfo{year}{2025}),
  ISSN \bibinfo{issn}{1476-4687}.

\bibitem[{\citenamefont{Li et~al.}(2025)\citenamefont{Li, Cai, Ren, Wang, Yang,
  Zhang, Wu, Chang, Wu, Jin et~al.}}]{Li2025}
\bibinfo{author}{\bibfnamefont{Y.}~\bibnamefont{Li}},
  \bibinfo{author}{\bibfnamefont{W.-Q.} \bibnamefont{Cai}},
  \bibinfo{author}{\bibfnamefont{J.-G.} \bibnamefont{Ren}},
  \bibinfo{author}{\bibfnamefont{C.-Z.} \bibnamefont{Wang}},
  \bibinfo{author}{\bibfnamefont{M.}~\bibnamefont{Yang}},
  \bibinfo{author}{\bibfnamefont{L.}~\bibnamefont{Zhang}},
  \bibinfo{author}{\bibfnamefont{H.-Y.} \bibnamefont{Wu}},
  \bibinfo{author}{\bibfnamefont{L.}~\bibnamefont{Chang}},
  \bibinfo{author}{\bibfnamefont{J.-C.} \bibnamefont{Wu}},
  \bibinfo{author}{\bibfnamefont{B.}~\bibnamefont{Jin}}, \bibnamefont{et~al.},
  \bibinfo{journal}{Nature} \textbf{\bibinfo{volume}{640}}, \bibinfo{pages}{47}
  (\bibinfo{year}{2025}), ISSN \bibinfo{issn}{1476-4687}.

\bibitem[{\citenamefont{Larocca et~al.}(2025)\citenamefont{Larocca, Thanasilp,
  Wang, Sharma, Biamonte, Coles, Cincio, McClean, Holmes, and
  Cerezo}}]{Larocca2025}
\bibinfo{author}{\bibfnamefont{M.}~\bibnamefont{Larocca}},
  \bibinfo{author}{\bibfnamefont{S.}~\bibnamefont{Thanasilp}},
  \bibinfo{author}{\bibfnamefont{S.}~\bibnamefont{Wang}},
  \bibinfo{author}{\bibfnamefont{K.}~\bibnamefont{Sharma}},
  \bibinfo{author}{\bibfnamefont{J.}~\bibnamefont{Biamonte}},
  \bibinfo{author}{\bibfnamefont{P.~J.} \bibnamefont{Coles}},
  \bibinfo{author}{\bibfnamefont{L.}~\bibnamefont{Cincio}},
  \bibinfo{author}{\bibfnamefont{J.~R.} \bibnamefont{McClean}},
  \bibinfo{author}{\bibfnamefont{Z.}~\bibnamefont{Holmes}}, \bibnamefont{and}
  \bibinfo{author}{\bibfnamefont{M.}~\bibnamefont{Cerezo}},
  \bibinfo{journal}{Nature Reviews Physics} \textbf{\bibinfo{volume}{7}},
  \bibinfo{pages}{174} (\bibinfo{year}{2025}), ISSN \bibinfo{issn}{2522-5820}.

\bibitem[{\citenamefont{Proctor et~al.}(2025)\citenamefont{Proctor, Young,
  Baczewski, and Blume-Kohout}}]{Proctor2025}
\bibinfo{author}{\bibfnamefont{T.}~\bibnamefont{Proctor}},
  \bibinfo{author}{\bibfnamefont{K.}~\bibnamefont{Young}},
  \bibinfo{author}{\bibfnamefont{A.~D.} \bibnamefont{Baczewski}},
  \bibnamefont{and}
  \bibinfo{author}{\bibfnamefont{R.}~\bibnamefont{Blume-Kohout}},
  \bibinfo{journal}{Nature Reviews Physics} \textbf{\bibinfo{volume}{7}},
  \bibinfo{pages}{105} (\bibinfo{year}{2025}), ISSN \bibinfo{issn}{2522-5820}.

\bibitem[{\citenamefont{Maring et~al.}(2024)\citenamefont{Maring, Fyrillas,
  Pont, Ivanov, Stepanov, Margaria, Hease, Pishchagin, Lemaître, Sagnes
  et~al.}}]{Maring2024}
\bibinfo{author}{\bibfnamefont{N.}~\bibnamefont{Maring}},
  \bibinfo{author}{\bibfnamefont{A.}~\bibnamefont{Fyrillas}},
  \bibinfo{author}{\bibfnamefont{M.}~\bibnamefont{Pont}},
  \bibinfo{author}{\bibfnamefont{E.}~\bibnamefont{Ivanov}},
  \bibinfo{author}{\bibfnamefont{P.}~\bibnamefont{Stepanov}},
  \bibinfo{author}{\bibfnamefont{N.}~\bibnamefont{Margaria}},
  \bibinfo{author}{\bibfnamefont{W.}~\bibnamefont{Hease}},
  \bibinfo{author}{\bibfnamefont{A.}~\bibnamefont{Pishchagin}},
  \bibinfo{author}{\bibfnamefont{A.}~\bibnamefont{Lemaître}},
  \bibinfo{author}{\bibfnamefont{I.}~\bibnamefont{Sagnes}},
  \bibnamefont{et~al.}, \bibinfo{journal}{Nature Photonics}
  \textbf{\bibinfo{volume}{18}}, \bibinfo{pages}{603} (\bibinfo{year}{2024}),
  ISSN \bibinfo{issn}{1749-4893}.

\bibitem[{\citenamefont{Cheng and Hall}(2015)}]{Cheng2015}
\bibinfo{author}{\bibfnamefont{S.}~\bibnamefont{Cheng}} \bibnamefont{and}
  \bibinfo{author}{\bibfnamefont{M.~J.~W.} \bibnamefont{Hall}},
  \bibinfo{journal}{Physical Review A} \textbf{\bibinfo{volume}{92}},
  \bibinfo{pages}{042101} (\bibinfo{year}{2015}), ISSN
  \bibinfo{issn}{1094-1622}.

\bibitem[{\citenamefont{Bemani et~al.}(2019)\citenamefont{Bemani, Roknizadeh,
  Motazedifard, Naderi, and Vitali}}]{Bemani2019}
\bibinfo{author}{\bibfnamefont{F.}~\bibnamefont{Bemani}},
  \bibinfo{author}{\bibfnamefont{R.}~\bibnamefont{Roknizadeh}},
  \bibinfo{author}{\bibfnamefont{A.}~\bibnamefont{Motazedifard}},
  \bibinfo{author}{\bibfnamefont{M.~H.} \bibnamefont{Naderi}},
  \bibnamefont{and} \bibinfo{author}{\bibfnamefont{D.}~\bibnamefont{Vitali}},
  \bibinfo{journal}{Physical Review A} \textbf{\bibinfo{volume}{99}},
  \bibinfo{pages}{063814} (\bibinfo{year}{2019}), ISSN
  \bibinfo{issn}{2469-9934}.

\bibitem[{\citenamefont{Emale et~al.}(2025)\citenamefont{Emale, Peng, Djorwé,
  Sarma, Abdourahimi, Abdel-Aty, Nisar, and Engo}}]{Emale.2025}
\bibinfo{author}{\bibfnamefont{K.}~\bibnamefont{Emale}},
  \bibinfo{author}{\bibfnamefont{J.-X.} \bibnamefont{Peng}},
  \bibinfo{author}{\bibfnamefont{P.}~\bibnamefont{Djorwé}},
  \bibinfo{author}{\bibfnamefont{A.}~\bibnamefont{Sarma}},
  \bibinfo{author}{\bibnamefont{Abdourahimi}},
  \bibinfo{author}{\bibfnamefont{A.-H.} \bibnamefont{Abdel-Aty}},
  \bibinfo{author}{\bibfnamefont{K.}~\bibnamefont{Nisar}}, \bibnamefont{and}
  \bibinfo{author}{\bibfnamefont{S.}~\bibnamefont{Engo}},
  \bibinfo{journal}{Physica B: Condensed Matter}
  \textbf{\bibinfo{volume}{701}}, \bibinfo{pages}{416919}
  (\bibinfo{year}{2025}), ISSN \bibinfo{issn}{0921-4526}.

\bibitem[{\citenamefont{Massembele et~al.}(2025)\citenamefont{Massembele,
  Djorwé, Emale, Peng, Abdel-Aty, and Nisar}}]{Rostand2025}
\bibinfo{author}{\bibfnamefont{D.}~\bibnamefont{Massembele}},
  \bibinfo{author}{\bibfnamefont{P.}~\bibnamefont{Djorwé}},
  \bibinfo{author}{\bibfnamefont{K.}~\bibnamefont{Emale}},
  \bibinfo{author}{\bibfnamefont{J.-X.} \bibnamefont{Peng}},
  \bibinfo{author}{\bibfnamefont{A.-H.} \bibnamefont{Abdel-Aty}},
  \bibnamefont{and} \bibinfo{author}{\bibfnamefont{K.}~\bibnamefont{Nisar}},
  \bibinfo{journal}{Physica B: Condensed Matter}
  \textbf{\bibinfo{volume}{697}}, \bibinfo{pages}{416689}
  (\bibinfo{year}{2025}), ISSN \bibinfo{issn}{0921-4526}.

\bibitem[{\citenamefont{Purdy et~al.}(2017)\citenamefont{Purdy, Grutter,
  Srinivasan, and Taylor}}]{Purdy2017}
\bibinfo{author}{\bibfnamefont{T.~P.} \bibnamefont{Purdy}},
  \bibinfo{author}{\bibfnamefont{K.~E.} \bibnamefont{Grutter}},
  \bibinfo{author}{\bibfnamefont{K.}~\bibnamefont{Srinivasan}},
  \bibnamefont{and} \bibinfo{author}{\bibfnamefont{J.~M.}
  \bibnamefont{Taylor}}, \bibinfo{journal}{Science}
  \textbf{\bibinfo{volume}{356}}, \bibinfo{pages}{1265} (\bibinfo{year}{2017}),
  ISSN \bibinfo{issn}{1095-9203}.

\bibitem[{\citenamefont{Zhu et~al.}(2017)\citenamefont{Zhu, Ma, Cao, Fei, and
  Vedral}}]{Zhu2017}
\bibinfo{author}{\bibfnamefont{H.}~\bibnamefont{Zhu}},
  \bibinfo{author}{\bibfnamefont{Z.}~\bibnamefont{Ma}},
  \bibinfo{author}{\bibfnamefont{Z.}~\bibnamefont{Cao}},
  \bibinfo{author}{\bibfnamefont{S.-M.} \bibnamefont{Fei}}, \bibnamefont{and}
  \bibinfo{author}{\bibfnamefont{V.}~\bibnamefont{Vedral}},
  \bibinfo{journal}{Physical Review A} \textbf{\bibinfo{volume}{96}},
  \bibinfo{pages}{032316} (\bibinfo{year}{2017}), ISSN
  \bibinfo{issn}{2469-9934}.

\bibitem[{\citenamefont{Dong et~al.}(2022)\citenamefont{Dong, Wei, Song, Wang,
  and Ye}}]{Dong2022}
\bibinfo{author}{\bibfnamefont{D.-D.} \bibnamefont{Dong}},
  \bibinfo{author}{\bibfnamefont{G.-B.} \bibnamefont{Wei}},
  \bibinfo{author}{\bibfnamefont{X.-K.} \bibnamefont{Song}},
  \bibinfo{author}{\bibfnamefont{D.}~\bibnamefont{Wang}}, \bibnamefont{and}
  \bibinfo{author}{\bibfnamefont{L.}~\bibnamefont{Ye}},
  \bibinfo{journal}{Physical Review A} \textbf{\bibinfo{volume}{106}},
  \bibinfo{pages}{042415} (\bibinfo{year}{2022}), ISSN
  \bibinfo{issn}{2469-9934}.

\bibitem[{\citenamefont{Zhang et~al.}(2016)\citenamefont{Zhang, Shao, Li, and
  Fan}}]{Zhang2016}
\bibinfo{author}{\bibfnamefont{Y.-R.} \bibnamefont{Zhang}},
  \bibinfo{author}{\bibfnamefont{L.-H.} \bibnamefont{Shao}},
  \bibinfo{author}{\bibfnamefont{Y.}~\bibnamefont{Li}}, \bibnamefont{and}
  \bibinfo{author}{\bibfnamefont{H.}~\bibnamefont{Fan}},
  \bibinfo{journal}{Physical Review A} \textbf{\bibinfo{volume}{93}},
  \bibinfo{pages}{012334} (\bibinfo{year}{2016}), ISSN
  \bibinfo{issn}{2469-9934}.

\bibitem[{\citenamefont{Xu}(2016)}]{Xu2016}
\bibinfo{author}{\bibfnamefont{J.}~\bibnamefont{Xu}},
  \bibinfo{journal}{Physical Review A} \textbf{\bibinfo{volume}{93}},
  \bibinfo{pages}{032111} (\bibinfo{year}{2016}), ISSN
  \bibinfo{issn}{2469-9934}.

\bibitem[{\citenamefont{Zheng et~al.}(2016)\citenamefont{Zheng, Xu, Yao, and
  Li}}]{Zheng2016}
\bibinfo{author}{\bibfnamefont{Q.}~\bibnamefont{Zheng}},
  \bibinfo{author}{\bibfnamefont{J.}~\bibnamefont{Xu}},
  \bibinfo{author}{\bibfnamefont{Y.}~\bibnamefont{Yao}}, \bibnamefont{and}
  \bibinfo{author}{\bibfnamefont{Y.}~\bibnamefont{Li}},
  \bibinfo{journal}{Physical Review A} \textbf{\bibinfo{volume}{94}},
  \bibinfo{pages}{052314} (\bibinfo{year}{2016}), ISSN
  \bibinfo{issn}{2469-9934}.

\bibitem[{\citenamefont{Peng et~al.}(2024)\citenamefont{Peng, Kundu, Liu,
  Rahman, Akhtar, and Asjad}}]{Peng2024}
\bibinfo{author}{\bibfnamefont{J.-X.} \bibnamefont{Peng}},
  \bibinfo{author}{\bibfnamefont{A.}~\bibnamefont{Kundu}},
  \bibinfo{author}{\bibfnamefont{Z.-X.} \bibnamefont{Liu}},
  \bibinfo{author}{\bibfnamefont{A.}~\bibnamefont{Rahman}},
  \bibinfo{author}{\bibfnamefont{N.}~\bibnamefont{Akhtar}}, \bibnamefont{and}
  \bibinfo{author}{\bibfnamefont{M.}~\bibnamefont{Asjad}},
  \bibinfo{journal}{Physical Review B} \textbf{\bibinfo{volume}{109}},
  \bibinfo{pages}{064412} (\bibinfo{year}{2024}), ISSN
  \bibinfo{issn}{2469-9969}.

\bibitem[{\citenamefont{Fröwis et~al.}(2018)\citenamefont{Fröwis, Yadin, and
  Gisin}}]{Frwis2018}
\bibinfo{author}{\bibfnamefont{F.}~\bibnamefont{Fröwis}},
  \bibinfo{author}{\bibfnamefont{B.}~\bibnamefont{Yadin}}, \bibnamefont{and}
  \bibinfo{author}{\bibfnamefont{N.}~\bibnamefont{Gisin}},
  \bibinfo{journal}{Physical Review A} \textbf{\bibinfo{volume}{97}},
  \bibinfo{pages}{042103} (\bibinfo{year}{2018}), ISSN
  \bibinfo{issn}{2469-9934}.

\bibitem[{\citenamefont{Donley et~al.}(2002)\citenamefont{Donley, Claussen,
  Thompson, and Wieman}}]{Donley2002}
\bibinfo{author}{\bibfnamefont{E.~A.} \bibnamefont{Donley}},
  \bibinfo{author}{\bibfnamefont{N.~R.} \bibnamefont{Claussen}},
  \bibinfo{author}{\bibfnamefont{S.~T.} \bibnamefont{Thompson}},
  \bibnamefont{and} \bibinfo{author}{\bibfnamefont{C.~E.}
  \bibnamefont{Wieman}}, \bibinfo{journal}{Nature}
  \textbf{\bibinfo{volume}{417}}, \bibinfo{pages}{529} (\bibinfo{year}{2002}),
  ISSN \bibinfo{issn}{1476-4687}.

\bibitem[{\citenamefont{Marquardt et~al.}(2007)\citenamefont{Marquardt,
  Andersen, Leuchs, Takeno, Yukawa, Yonezawa, and Furusawa}}]{Marquardt2007}
\bibinfo{author}{\bibfnamefont{C.}~\bibnamefont{Marquardt}},
  \bibinfo{author}{\bibfnamefont{U.~L.} \bibnamefont{Andersen}},
  \bibinfo{author}{\bibfnamefont{G.}~\bibnamefont{Leuchs}},
  \bibinfo{author}{\bibfnamefont{Y.}~\bibnamefont{Takeno}},
  \bibinfo{author}{\bibfnamefont{M.}~\bibnamefont{Yukawa}},
  \bibinfo{author}{\bibfnamefont{H.}~\bibnamefont{Yonezawa}}, \bibnamefont{and}
  \bibinfo{author}{\bibfnamefont{A.}~\bibnamefont{Furusawa}},
  \bibinfo{journal}{Physical Review A} \textbf{\bibinfo{volume}{76}},
  \bibinfo{pages}{030101(R)} (\bibinfo{year}{2007}), ISSN
  \bibinfo{issn}{1094-1622}.

\bibitem[{\citenamefont{Kang et~al.}(2021)\citenamefont{Kang, Han, Wang, Liu,
  Hao, and Su}}]{Kang2021}
\bibinfo{author}{\bibfnamefont{H.}~\bibnamefont{Kang}},
  \bibinfo{author}{\bibfnamefont{D.}~\bibnamefont{Han}},
  \bibinfo{author}{\bibfnamefont{N.}~\bibnamefont{Wang}},
  \bibinfo{author}{\bibfnamefont{Y.}~\bibnamefont{Liu}},
  \bibinfo{author}{\bibfnamefont{S.}~\bibnamefont{Hao}}, \bibnamefont{and}
  \bibinfo{author}{\bibfnamefont{X.}~\bibnamefont{Su}},
  \bibinfo{journal}{Photonics Research} \textbf{\bibinfo{volume}{9}},
  \bibinfo{pages}{1330} (\bibinfo{year}{2021}), ISSN \bibinfo{issn}{2327-9125}.

\bibitem[{\citenamefont{Zhai et~al.}(2023)\citenamefont{Zhai, Du, and
  Guo}}]{Zhai2023}
\bibinfo{author}{\bibfnamefont{L.-l.} \bibnamefont{Zhai}},
  \bibinfo{author}{\bibfnamefont{H.-J.} \bibnamefont{Du}}, \bibnamefont{and}
  \bibinfo{author}{\bibfnamefont{J.-L.} \bibnamefont{Guo}},
  \bibinfo{journal}{Quantum Information Processing}
  \textbf{\bibinfo{volume}{22}}, \bibinfo{pages}{211} (\bibinfo{year}{2023}),
  ISSN \bibinfo{issn}{1573-1332}.

\bibitem[{\citenamefont{Wu et~al.}(2023)\citenamefont{Wu, Bai, Li, Yu, and
  Zhang}}]{Wu2023}
\bibinfo{author}{\bibfnamefont{S.-X.} \bibnamefont{Wu}},
  \bibinfo{author}{\bibfnamefont{C.-H.} \bibnamefont{Bai}},
  \bibinfo{author}{\bibfnamefont{G.}~\bibnamefont{Li}},
  \bibinfo{author}{\bibfnamefont{C.-s.} \bibnamefont{Yu}}, \bibnamefont{and}
  \bibinfo{author}{\bibfnamefont{T.}~\bibnamefont{Zhang}},
  \bibinfo{journal}{Optics Express} \textbf{\bibinfo{volume}{32}},
  \bibinfo{pages}{260} (\bibinfo{year}{2023}), ISSN \bibinfo{issn}{1094-4087}.

\bibitem[{\citenamefont{Djorwé
  et~al.}(2024{\natexlab{a}})\citenamefont{Djorwé, Abdel-Aty, Nisar, and
  Engo}}]{Rost2024}
\bibinfo{author}{\bibfnamefont{P.}~\bibnamefont{Djorwé}},
  \bibinfo{author}{\bibfnamefont{A.-H.} \bibnamefont{Abdel-Aty}},
  \bibinfo{author}{\bibfnamefont{K.}~\bibnamefont{Nisar}}, \bibnamefont{and}
  \bibinfo{author}{\bibfnamefont{S.}~\bibnamefont{Engo}},
  \bibinfo{journal}{Optik} \textbf{\bibinfo{volume}{319}},
  \bibinfo{pages}{172097} (\bibinfo{year}{2024}{\natexlab{a}}), ISSN
  \bibinfo{issn}{0030-4026}.

\bibitem[{\citenamefont{Fan et~al.}(2024)\citenamefont{Fan, Zuo, Li, and
  Li}}]{Fan2024}
\bibinfo{author}{\bibfnamefont{Z.}~\bibnamefont{Fan}},
  \bibinfo{author}{\bibfnamefont{X.}~\bibnamefont{Zuo}},
  \bibinfo{author}{\bibfnamefont{H.-T.} \bibnamefont{Li}}, \bibnamefont{and}
  \bibinfo{author}{\bibfnamefont{J.}~\bibnamefont{Li}},
  \bibinfo{journal}{Fundamental Research} p. \bibinfo{pages}{1958}
  (\bibinfo{year}{2024}), ISSN \bibinfo{issn}{2667-3258}.

\bibitem[{\citenamefont{Hussain et~al.}(2022)\citenamefont{Hussain, Qamar, and
  Irfan}}]{Hussain2022}
\bibinfo{author}{\bibfnamefont{B.}~\bibnamefont{Hussain}},
  \bibinfo{author}{\bibfnamefont{S.}~\bibnamefont{Qamar}}, \bibnamefont{and}
  \bibinfo{author}{\bibfnamefont{M.}~\bibnamefont{Irfan}},
  \bibinfo{journal}{Physical Review A: Atomic, Molecular, and Optical Physics}
  \textbf{\bibinfo{volume}{105}}, \bibinfo{pages}{063704}
  (\bibinfo{year}{2022}),
  \urlprefix\url{https://link.aps.org/doi/10.1103/PhysRevA.105.063704}.

\bibitem[{\citenamefont{Pavlovich et~al.}(2025)\citenamefont{Pavlovich, Rakich,
  and Puri}}]{Puri.2025}
\bibinfo{author}{\bibfnamefont{M.}~\bibnamefont{Pavlovich}},
  \bibinfo{author}{\bibfnamefont{P.}~\bibnamefont{Rakich}}, \bibnamefont{and}
  \bibinfo{author}{\bibfnamefont{S.}~\bibnamefont{Puri}}
  (\bibinfo{year}{2025}).

\bibitem[{\citenamefont{Li et~al.}(2021)\citenamefont{Li, Ou, Lei, and
  Liu}}]{Li2021}
\bibinfo{author}{\bibfnamefont{B.-B.} \bibnamefont{Li}},
  \bibinfo{author}{\bibfnamefont{L.}~\bibnamefont{Ou}},
  \bibinfo{author}{\bibfnamefont{Y.}~\bibnamefont{Lei}}, \bibnamefont{and}
  \bibinfo{author}{\bibfnamefont{Y.-C.} \bibnamefont{Liu}},
  \bibinfo{journal}{Nanophotonics} \textbf{\bibinfo{volume}{10}},
  \bibinfo{pages}{2799} (\bibinfo{year}{2021}), ISSN \bibinfo{issn}{2192-8614}.

\bibitem[{\citenamefont{Djorwé
  et~al.}(2024{\natexlab{b}})\citenamefont{Djorwé, Asjad, Pennec, Dutykh, and
  Djafari-Rouhani}}]{Dj2024}
\bibinfo{author}{\bibfnamefont{P.}~\bibnamefont{Djorwé}},
  \bibinfo{author}{\bibfnamefont{M.}~\bibnamefont{Asjad}},
  \bibinfo{author}{\bibfnamefont{Y.}~\bibnamefont{Pennec}},
  \bibinfo{author}{\bibfnamefont{D.}~\bibnamefont{Dutykh}}, \bibnamefont{and}
  \bibinfo{author}{\bibfnamefont{B.}~\bibnamefont{Djafari-Rouhani}},
  \bibinfo{journal}{Physical Review Research} \textbf{\bibinfo{volume}{6}},
  \bibinfo{pages}{033284} (\bibinfo{year}{2024}{\natexlab{b}}), ISSN
  \bibinfo{issn}{2643-1564}.

\bibitem[{\citenamefont{Xia et~al.}(2023)\citenamefont{Xia, Agrawal, Pluchar,
  Brady, Liu, Zhuang, Wilson, and Zhang}}]{Xia2023}
\bibinfo{author}{\bibfnamefont{Y.}~\bibnamefont{Xia}},
  \bibinfo{author}{\bibfnamefont{A.~R.} \bibnamefont{Agrawal}},
  \bibinfo{author}{\bibfnamefont{C.~M.} \bibnamefont{Pluchar}},
  \bibinfo{author}{\bibfnamefont{A.~J.} \bibnamefont{Brady}},
  \bibinfo{author}{\bibfnamefont{Z.}~\bibnamefont{Liu}},
  \bibinfo{author}{\bibfnamefont{Q.}~\bibnamefont{Zhuang}},
  \bibinfo{author}{\bibfnamefont{D.~J.} \bibnamefont{Wilson}},
  \bibnamefont{and} \bibinfo{author}{\bibfnamefont{Z.}~\bibnamefont{Zhang}},
  \bibinfo{journal}{Nature Photonics} \textbf{\bibinfo{volume}{17}},
  \bibinfo{pages}{470} (\bibinfo{year}{2023}), ISSN \bibinfo{issn}{1749-4893}.

\bibitem[{\citenamefont{Tang et~al.}(2023)\citenamefont{Tang, Qin, Liu, Wang,
  Cui, Su, Yan, and Chen}}]{Tang2023}
\bibinfo{author}{\bibfnamefont{S.-B.} \bibnamefont{Tang}},
  \bibinfo{author}{\bibfnamefont{H.}~\bibnamefont{Qin}},
  \bibinfo{author}{\bibfnamefont{B.-B.} \bibnamefont{Liu}},
  \bibinfo{author}{\bibfnamefont{D.-Y.} \bibnamefont{Wang}},
  \bibinfo{author}{\bibfnamefont{K.}~\bibnamefont{Cui}},
  \bibinfo{author}{\bibfnamefont{S.-L.} \bibnamefont{Su}},
  \bibinfo{author}{\bibfnamefont{L.-L.} \bibnamefont{Yan}}, \bibnamefont{and}
  \bibinfo{author}{\bibfnamefont{G.}~\bibnamefont{Chen}},
  \bibinfo{journal}{Physical Review A} \textbf{\bibinfo{volume}{108}},
  \bibinfo{pages}{053514} (\bibinfo{year}{2023}), ISSN
  \bibinfo{issn}{2469-9934}.

\bibitem[{\citenamefont{Djorwe et~al.}(2019)\citenamefont{Djorwe, Pennec, and
  Djafari-Rouhani}}]{Djorwe2019}
\bibinfo{author}{\bibfnamefont{P.}~\bibnamefont{Djorwe}},
  \bibinfo{author}{\bibfnamefont{Y.}~\bibnamefont{Pennec}}, \bibnamefont{and}
  \bibinfo{author}{\bibfnamefont{B.}~\bibnamefont{Djafari-Rouhani}},
  \bibinfo{journal}{Physical Review Applied} \textbf{\bibinfo{volume}{12}},
  \bibinfo{pages}{024002} (\bibinfo{year}{2019}), ISSN
  \bibinfo{issn}{2331-7019}.

\bibitem[{\citenamefont{Pezzè et~al.}(2018)\citenamefont{Pezzè, Smerzi,
  Oberthaler, Schmied, and Treutlein}}]{Pezz2018}
\bibinfo{author}{\bibfnamefont{L.}~\bibnamefont{Pezzè}},
  \bibinfo{author}{\bibfnamefont{A.}~\bibnamefont{Smerzi}},
  \bibinfo{author}{\bibfnamefont{M.~K.} \bibnamefont{Oberthaler}},
  \bibinfo{author}{\bibfnamefont{R.}~\bibnamefont{Schmied}}, \bibnamefont{and}
  \bibinfo{author}{\bibfnamefont{P.}~\bibnamefont{Treutlein}},
  \bibinfo{journal}{Reviews of Modern Physics} \textbf{\bibinfo{volume}{90}},
  \bibinfo{pages}{035005} (\bibinfo{year}{2018}), ISSN
  \bibinfo{issn}{1539-0756}.

\bibitem[{\citenamefont{Bai et~al.}(2019)\citenamefont{Bai, Peng, Luo, and
  An}}]{Bai2019}
\bibinfo{author}{\bibfnamefont{K.}~\bibnamefont{Bai}},
  \bibinfo{author}{\bibfnamefont{Z.}~\bibnamefont{Peng}},
  \bibinfo{author}{\bibfnamefont{H.-G.} \bibnamefont{Luo}}, \bibnamefont{and}
  \bibinfo{author}{\bibfnamefont{J.-H.} \bibnamefont{An}},
  \bibinfo{journal}{Physical Review Letters} \textbf{\bibinfo{volume}{123}},
  \bibinfo{pages}{040402} (\bibinfo{year}{2019}), ISSN
  \bibinfo{issn}{1079-7114}.

\bibitem[{\citenamefont{Lai et~al.}(2022{\natexlab{a}})\citenamefont{Lai, Chen,
  Qin, Miranowicz, and Nori}}]{Lai2022}
\bibinfo{author}{\bibfnamefont{D.-G.} \bibnamefont{Lai}},
  \bibinfo{author}{\bibfnamefont{Y.-H.} \bibnamefont{Chen}},
  \bibinfo{author}{\bibfnamefont{W.}~\bibnamefont{Qin}},
  \bibinfo{author}{\bibfnamefont{A.}~\bibnamefont{Miranowicz}},
  \bibnamefont{and} \bibinfo{author}{\bibfnamefont{F.}~\bibnamefont{Nori}},
  \bibinfo{journal}{Physical Review Research} \textbf{\bibinfo{volume}{4}},
  \bibinfo{pages}{033112} (\bibinfo{year}{2022}{\natexlab{a}}), ISSN
  \bibinfo{issn}{2643-1564}.

\bibitem[{\citenamefont{Lai et~al.}(2022{\natexlab{b}})\citenamefont{Lai, Liao,
  Miranowicz, and Nori}}]{Nori2022}
\bibinfo{author}{\bibfnamefont{D.-G.} \bibnamefont{Lai}},
  \bibinfo{author}{\bibfnamefont{J.-Q.} \bibnamefont{Liao}},
  \bibinfo{author}{\bibfnamefont{A.}~\bibnamefont{Miranowicz}},
  \bibnamefont{and} \bibinfo{author}{\bibfnamefont{F.}~\bibnamefont{Nori}},
  \bibinfo{journal}{Physical Review Letters} \textbf{\bibinfo{volume}{129}},
  \bibinfo{pages}{063602} (\bibinfo{year}{2022}{\natexlab{b}}), ISSN
  \bibinfo{issn}{1079-7114}.

\bibitem[{\citenamefont{Qiu et~al.}(2022)\citenamefont{Qiu, Cheng, Chen, Lan,
  and Nie}}]{Qiu2022}
\bibinfo{author}{\bibfnamefont{W.}~\bibnamefont{Qiu}},
  \bibinfo{author}{\bibfnamefont{X.}~\bibnamefont{Cheng}},
  \bibinfo{author}{\bibfnamefont{A.}~\bibnamefont{Chen}},
  \bibinfo{author}{\bibfnamefont{Y.}~\bibnamefont{Lan}}, \bibnamefont{and}
  \bibinfo{author}{\bibfnamefont{W.}~\bibnamefont{Nie}},
  \bibinfo{journal}{Physical Review A} \textbf{\bibinfo{volume}{105}},
  \bibinfo{pages}{063718} (\bibinfo{year}{2022}), ISSN
  \bibinfo{issn}{2469-9934}.

\bibitem[{\citenamefont{Liu et~al.}(2025)\citenamefont{Liu, Jiao, Yin, Yu,
  Wang, and Jing}}]{Liu2025}
\bibinfo{author}{\bibfnamefont{J.-X.} \bibnamefont{Liu}},
  \bibinfo{author}{\bibfnamefont{Y.-F.} \bibnamefont{Jiao}},
  \bibinfo{author}{\bibfnamefont{B.}~\bibnamefont{Yin}},
  \bibinfo{author}{\bibfnamefont{H.-Y.} \bibnamefont{Yu}},
  \bibinfo{author}{\bibfnamefont{R.-C.} \bibnamefont{Wang}}, \bibnamefont{and}
  \bibinfo{author}{\bibfnamefont{H.}~\bibnamefont{Jing}},
  \bibinfo{journal}{Physical Review A} \textbf{\bibinfo{volume}{112}},
  \bibinfo{pages}{013535} (\bibinfo{year}{2025}), ISSN
  \bibinfo{issn}{2469-9934}.

\bibitem[{\citenamefont{Jiao et~al.}(2025)\citenamefont{Jiao, Wang, Wang, Tang,
  Wang, Zuo, Bao, Kuang, and Jing}}]{Jiao2025}
\bibinfo{author}{\bibfnamefont{Y.-F.} \bibnamefont{Jiao}},
  \bibinfo{author}{\bibfnamefont{J.}~\bibnamefont{Wang}},
  \bibinfo{author}{\bibfnamefont{D.-Y.} \bibnamefont{Wang}},
  \bibinfo{author}{\bibfnamefont{L.}~\bibnamefont{Tang}},
  \bibinfo{author}{\bibfnamefont{Y.}~\bibnamefont{Wang}},
  \bibinfo{author}{\bibfnamefont{Y.-L.} \bibnamefont{Zuo}},
  \bibinfo{author}{\bibfnamefont{W.-S.} \bibnamefont{Bao}},
  \bibinfo{author}{\bibfnamefont{L.-M.} \bibnamefont{Kuang}}, \bibnamefont{and}
  \bibinfo{author}{\bibfnamefont{H.}~\bibnamefont{Jing}},
  \bibinfo{journal}{Physical Review A} \textbf{\bibinfo{volume}{112}},
  \bibinfo{pages}{012421} (\bibinfo{year}{2025}), ISSN
  \bibinfo{issn}{2469-9934}.

\bibitem[{\citenamefont{Ge et~al.}(2025)\citenamefont{Ge, Yu, Wu, Han, Wang,
  and Zhang}}]{Ge2025}
\bibinfo{author}{\bibfnamefont{P.-C.} \bibnamefont{Ge}},
  \bibinfo{author}{\bibfnamefont{Y.}~\bibnamefont{Yu}},
  \bibinfo{author}{\bibfnamefont{H.-T.} \bibnamefont{Wu}},
  \bibinfo{author}{\bibfnamefont{X.}~\bibnamefont{Han}},
  \bibinfo{author}{\bibfnamefont{H.-F.} \bibnamefont{Wang}}, \bibnamefont{and}
  \bibinfo{author}{\bibfnamefont{S.}~\bibnamefont{Zhang}},
  \bibinfo{journal}{Scientific Reports} \textbf{\bibinfo{volume}{15}},
  \bibinfo{pages}{7937} (\bibinfo{year}{2025}), ISSN \bibinfo{issn}{2045-2322}.

\bibitem[{\citenamefont{Lu et~al.}(2025)\citenamefont{Lu, Li, Chen, Wang, Xiao,
  and Jing}}]{Lu2025}
\bibinfo{author}{\bibfnamefont{T.-X.} \bibnamefont{Lu}},
  \bibinfo{author}{\bibfnamefont{Z.-S.} \bibnamefont{Li}},
  \bibinfo{author}{\bibfnamefont{L.-S.} \bibnamefont{Chen}},
  \bibinfo{author}{\bibfnamefont{Y.}~\bibnamefont{Wang}},
  \bibinfo{author}{\bibfnamefont{X.}~\bibnamefont{Xiao}}, \bibnamefont{and}
  \bibinfo{author}{\bibfnamefont{H.}~\bibnamefont{Jing}},
  \bibinfo{journal}{Physical Review A} \textbf{\bibinfo{volume}{111}},
  \bibinfo{pages}{013713} (\bibinfo{year}{2025}), ISSN
  \bibinfo{issn}{2469-9934}.

\bibitem[{\citenamefont{Cai et~al.}(2025)\citenamefont{Cai, Fan, Tang, Chen,
  and Deng}}]{Cai2025}
\bibinfo{author}{\bibfnamefont{Q.}~\bibnamefont{Cai}},
  \bibinfo{author}{\bibfnamefont{B.}~\bibnamefont{Fan}},
  \bibinfo{author}{\bibfnamefont{J.-D.} \bibnamefont{Tang}},
  \bibinfo{author}{\bibfnamefont{H.}~\bibnamefont{Chen}}, \bibnamefont{and}
  \bibinfo{author}{\bibfnamefont{G.}~\bibnamefont{Deng}},
  \bibinfo{journal}{Optics and Laser Technology}
  \textbf{\bibinfo{volume}{182}}, \bibinfo{pages}{112100}
  (\bibinfo{year}{2025}), ISSN \bibinfo{issn}{0030-3992}.

\bibitem[{\citenamefont{Chen et~al.}(2023)\citenamefont{Chen, Fan, Xiong, Wang,
  and Ye}}]{Chen2023}
\bibinfo{author}{\bibfnamefont{J.}~\bibnamefont{Chen}},
  \bibinfo{author}{\bibfnamefont{X.-G.} \bibnamefont{Fan}},
  \bibinfo{author}{\bibfnamefont{W.}~\bibnamefont{Xiong}},
  \bibinfo{author}{\bibfnamefont{D.}~\bibnamefont{Wang}}, \bibnamefont{and}
  \bibinfo{author}{\bibfnamefont{L.}~\bibnamefont{Ye}},
  \bibinfo{journal}{Physical Review B} \textbf{\bibinfo{volume}{108}},
  \bibinfo{pages}{024105} (\bibinfo{year}{2023}), ISSN
  \bibinfo{issn}{2469-9969}.

\bibitem[{\citenamefont{Bai et~al.}(2022)\citenamefont{Bai, Fang, Liu, Li, Wan,
  and Xiao}}]{Bai2022}
\bibinfo{author}{\bibfnamefont{K.}~\bibnamefont{Bai}},
  \bibinfo{author}{\bibfnamefont{L.}~\bibnamefont{Fang}},
  \bibinfo{author}{\bibfnamefont{T.-R.} \bibnamefont{Liu}},
  \bibinfo{author}{\bibfnamefont{J.-Z.} \bibnamefont{Li}},
  \bibinfo{author}{\bibfnamefont{D.}~\bibnamefont{Wan}}, \bibnamefont{and}
  \bibinfo{author}{\bibfnamefont{M.}~\bibnamefont{Xiao}},
  \bibinfo{journal}{National Science Review} \textbf{\bibinfo{volume}{10}},
  \bibinfo{pages}{nwac259} (\bibinfo{year}{2022}), ISSN
  \bibinfo{issn}{2053-714X}.

\bibitem[{\citenamefont{Bai et~al.}(2023)\citenamefont{Bai, Li, Liu, Fang, Wan,
  and Xiao}}]{Bai2023}
\bibinfo{author}{\bibfnamefont{K.}~\bibnamefont{Bai}},
  \bibinfo{author}{\bibfnamefont{J.-Z.} \bibnamefont{Li}},
  \bibinfo{author}{\bibfnamefont{T.-R.} \bibnamefont{Liu}},
  \bibinfo{author}{\bibfnamefont{L.}~\bibnamefont{Fang}},
  \bibinfo{author}{\bibfnamefont{D.}~\bibnamefont{Wan}}, \bibnamefont{and}
  \bibinfo{author}{\bibfnamefont{M.}~\bibnamefont{Xiao}},
  \bibinfo{journal}{Physical Review Letters} \textbf{\bibinfo{volume}{130}},
  \bibinfo{pages}{266901} (\bibinfo{year}{2023}), ISSN
  \bibinfo{issn}{1079-7114}.

\bibitem[{\citenamefont{Bai et~al.}(2024)\citenamefont{Bai, Liu, Fang, Li, Lin,
  Wan, and Xiao}}]{Bai2024a}
\bibinfo{author}{\bibfnamefont{K.}~\bibnamefont{Bai}},
  \bibinfo{author}{\bibfnamefont{T.-R.} \bibnamefont{Liu}},
  \bibinfo{author}{\bibfnamefont{L.}~\bibnamefont{Fang}},
  \bibinfo{author}{\bibfnamefont{J.-Z.} \bibnamefont{Li}},
  \bibinfo{author}{\bibfnamefont{C.}~\bibnamefont{Lin}},
  \bibinfo{author}{\bibfnamefont{D.}~\bibnamefont{Wan}}, \bibnamefont{and}
  \bibinfo{author}{\bibfnamefont{M.}~\bibnamefont{Xiao}},
  \bibinfo{journal}{Physical Review Letters} \textbf{\bibinfo{volume}{132}},
  \bibinfo{pages}{073802} (\bibinfo{year}{2024}), ISSN
  \bibinfo{issn}{1079-7114}.

\bibitem[{\citenamefont{Schmidt et~al.}(2015)\citenamefont{Schmidt, Kessler,
  Peano, Painter, and Marquardt}}]{Schmidt2015}
\bibinfo{author}{\bibfnamefont{M.}~\bibnamefont{Schmidt}},
  \bibinfo{author}{\bibfnamefont{S.}~\bibnamefont{Kessler}},
  \bibinfo{author}{\bibfnamefont{V.}~\bibnamefont{Peano}},
  \bibinfo{author}{\bibfnamefont{O.}~\bibnamefont{Painter}}, \bibnamefont{and}
  \bibinfo{author}{\bibfnamefont{F.}~\bibnamefont{Marquardt}},
  \bibinfo{journal}{Optica} \textbf{\bibinfo{volume}{2}}, \bibinfo{pages}{635}
  (\bibinfo{year}{2015}), ISSN \bibinfo{issn}{2334-2536}.

\bibitem[{\citenamefont{Fang et~al.}(2017)\citenamefont{Fang, Luo, Metelmann,
  Matheny, Marquardt, Clerk, and Painter}}]{Fang2017}
\bibinfo{author}{\bibfnamefont{K.}~\bibnamefont{Fang}},
  \bibinfo{author}{\bibfnamefont{J.}~\bibnamefont{Luo}},
  \bibinfo{author}{\bibfnamefont{A.}~\bibnamefont{Metelmann}},
  \bibinfo{author}{\bibfnamefont{M.~H.} \bibnamefont{Matheny}},
  \bibinfo{author}{\bibfnamefont{F.}~\bibnamefont{Marquardt}},
  \bibinfo{author}{\bibfnamefont{A.~A.} \bibnamefont{Clerk}}, \bibnamefont{and}
  \bibinfo{author}{\bibfnamefont{O.}~\bibnamefont{Painter}},
  \bibinfo{journal}{Nature Physics} \textbf{\bibinfo{volume}{13}},
  \bibinfo{pages}{465} (\bibinfo{year}{2017}), ISSN \bibinfo{issn}{1745-2481}.

\bibitem[{\citenamefont{Brendel et~al.}(2017)\citenamefont{Brendel, Peano,
  Painter, and Marquardt}}]{Brendel2017}
\bibinfo{author}{\bibfnamefont{C.}~\bibnamefont{Brendel}},
  \bibinfo{author}{\bibfnamefont{V.}~\bibnamefont{Peano}},
  \bibinfo{author}{\bibfnamefont{O.~J.} \bibnamefont{Painter}},
  \bibnamefont{and}
  \bibinfo{author}{\bibfnamefont{F.}~\bibnamefont{Marquardt}},
  \bibinfo{journal}{Proceedings of the National Academy of Sciences}
  \textbf{\bibinfo{volume}{114}}, \bibinfo{pages}{E3390}
  (\bibinfo{year}{2017}), ISSN \bibinfo{issn}{1091-6490}.

\bibitem[{\citenamefont{Mathew et~al.}(2020)\citenamefont{Mathew, Pino, and
  Verhagen}}]{Mathew2020}
\bibinfo{author}{\bibfnamefont{J.~P.} \bibnamefont{Mathew}},
  \bibinfo{author}{\bibfnamefont{J.~d.} \bibnamefont{Pino}}, \bibnamefont{and}
  \bibinfo{author}{\bibfnamefont{E.}~\bibnamefont{Verhagen}},
  \bibinfo{journal}{Nature Nanotechnology} \textbf{\bibinfo{volume}{15}},
  \bibinfo{pages}{198} (\bibinfo{year}{2020}), ISSN \bibinfo{issn}{1748-3395}.

\bibitem[{\citenamefont{Chen et~al.}(2021)\citenamefont{Chen, Zhang, Shen, Zou,
  Guo, and Dong}}]{Chen2021}
\bibinfo{author}{\bibfnamefont{Y.}~\bibnamefont{Chen}},
  \bibinfo{author}{\bibfnamefont{Y.-L.} \bibnamefont{Zhang}},
  \bibinfo{author}{\bibfnamefont{Z.}~\bibnamefont{Shen}},
  \bibinfo{author}{\bibfnamefont{C.-L.} \bibnamefont{Zou}},
  \bibinfo{author}{\bibfnamefont{G.-C.} \bibnamefont{Guo}}, \bibnamefont{and}
  \bibinfo{author}{\bibfnamefont{C.-H.} \bibnamefont{Dong}},
  \bibinfo{journal}{Physical Review Letters} \textbf{\bibinfo{volume}{126}},
  \bibinfo{pages}{123603} (\bibinfo{year}{2021}), ISSN
  \bibinfo{issn}{1079-7114}.

\bibitem[{\citenamefont{Slim et~al.}(2025)\citenamefont{Slim, del Pino, and
  Verhagen}}]{Slim2025}
\bibinfo{author}{\bibfnamefont{J.~J.} \bibnamefont{Slim}},
  \bibinfo{author}{\bibfnamefont{J.}~\bibnamefont{del Pino}}, \bibnamefont{and}
  \bibinfo{author}{\bibfnamefont{E.}~\bibnamefont{Verhagen}},
  \bibinfo{journal}{Nature Communications} \textbf{\bibinfo{volume}{16}},
  \bibinfo{pages}{7471} (\bibinfo{year}{2025}), ISSN \bibinfo{issn}{2041-1723}.

\bibitem[{\citenamefont{Hassan et~al.}(2015)\citenamefont{Hassan, Hodaei, Miri,
  Khajavikhan, and Christodoulides}}]{Hassan2015}
\bibinfo{author}{\bibfnamefont{A.~U.} \bibnamefont{Hassan}},
  \bibinfo{author}{\bibfnamefont{H.}~\bibnamefont{Hodaei}},
  \bibinfo{author}{\bibfnamefont{M.-A.} \bibnamefont{Miri}},
  \bibinfo{author}{\bibfnamefont{M.}~\bibnamefont{Khajavikhan}},
  \bibnamefont{and} \bibinfo{author}{\bibfnamefont{D.~N.}
  \bibnamefont{Christodoulides}}, \bibinfo{journal}{Physical Review A}
  \textbf{\bibinfo{volume}{92}}, \bibinfo{pages}{063807}
  (\bibinfo{year}{2015}), ISSN \bibinfo{issn}{1094-1622}.

\bibitem[{\citenamefont{DeJesus and Kaufman}(1987)}]{DeJesus}
\bibinfo{author}{\bibfnamefont{E.~X.} \bibnamefont{DeJesus}} \bibnamefont{and}
  \bibinfo{author}{\bibfnamefont{C.}~\bibnamefont{Kaufman}},
  \bibinfo{journal}{Physical Review A} \textbf{\bibinfo{volume}{35}},
  \bibinfo{pages}{5288} (\bibinfo{year}{1987}), ISSN \bibinfo{issn}{0556-2791}.

\bibitem[{\citenamefont{Adesso and Illuminati}(2006)}]{Adesso2006}
\bibinfo{author}{\bibfnamefont{G.}~\bibnamefont{Adesso}} \bibnamefont{and}
  \bibinfo{author}{\bibfnamefont{F.}~\bibnamefont{Illuminati}},
  \bibinfo{journal}{New Journal of Physics} \textbf{\bibinfo{volume}{8}},
  \bibinfo{pages}{15} (\bibinfo{year}{2006}), ISSN \bibinfo{issn}{1367-2630}.

\bibitem[{\citenamefont{Adesso and Illuminati}(2007)}]{Adesso2007}
\bibinfo{author}{\bibfnamefont{G.}~\bibnamefont{Adesso}} \bibnamefont{and}
  \bibinfo{author}{\bibfnamefont{F.}~\bibnamefont{Illuminati}},
  \bibinfo{journal}{Journal of Physics A: Mathematical and Theoretical}
  \textbf{\bibinfo{volume}{40}}, \bibinfo{pages}{7821} (\bibinfo{year}{2007}),
  ISSN \bibinfo{issn}{1751-8121}.

\end{thebibliography}

\end{document}